\documentclass[aps,prx,twocolumn,preprintnumbers,amsmath,amssymb]{revtex4-2}
\usepackage{dcolumn}
\usepackage{bm}
\usepackage{graphicx}
\usepackage{amsmath,amsfonts,amssymb,amsthm,amstext,amsbsy,amsopn,amscd,amsxtra}
\usepackage{dsfont}
\usepackage{physics} 
\usepackage{upgreek} 
\usepackage[colorlinks=true,allcolors=blue]{hyperref}
\usepackage{xcolor}
\usepackage{url}
\usepackage{siunitx}
\usepackage{multirow}
\usepackage{lineno}

\def\degree{^\mathrm{o}}
\def\dL{\text{L}}
\def\dTOF{\text{TOF}}

\everymath{\displaystyle}

\begin{document}

\title{Emergence of a tunable crystalline order in a Floquet-Bloch system from a parametric instability}

\author{N. Dupont$^1$, L. Gabardos$^1$, F. Arrouas$^1$, G. Chatelain$^1$, M. Arnal$^1$, J. Billy$^1$, P. Schlagheck$^2$, B. Peaudecerf$^1$, D. Gu\'ery-Odelin$^1$}
\email{dgo@irsamc.ups-tlse.fr}
\affiliation{
$^1$ Laboratoire Collisions Agr\'egats R\'eactivit\'e, UMR 5589, FERMI, UT3, Universit\'e de Toulouse, CNRS,\\
118 Route de Narbonne, 31062 Toulouse CEDEX 09, France \\
$^2$ CESAM Research Unit, University of Li\`ege,\\ 
4000 Li\`ege, Belgium
}
\date{\today}

\begin{abstract}

Parametric instabilities in interacting systems can lead to the appearance of new structures or patterns. In quantum gases, two-body interactions are responsible for a variety of instabilities that depend on the characteristics of both trapping and interactions. We report on the Floquet engineering of such instabilities, on a Bose-Einstein condensate held in a time-modulated optical lattice. The modulation triggers a destabilization of the condensate into a state exhibiting a density modulation with a new spatial periodicity. This new crystal-like order directly depends on the modulation parameters: the interplay between the Floquet spectrum and interactions generates narrow and adjustable instability regions, leading to the growth, from quantum or thermal fluctuations, of modes with a density modulation non commensurate with the lattice spacing. This study demonstrates the production of metastable exotic states of matter through Floquet engineering, and paves the way for further studies of dissipation in the resulting phase, and of similar phenomena in other geometries.

\end{abstract}

\maketitle

\section*{Introduction}

The introduction of interactions in wave theory leads, through non-linearities, to a rich phenomenology. In quantum gases it is at the root of the modification of the equilibrium momentum distribution through the production of momentum-correlated pairs~\cite{Lopes17,Xu06,Cayla20,Tenart21}. It can also lead to instabilities that are responsible for a new structuration of the gas, and the appearance of patterns. Such pattern formation may occur in static systems or through parameter quenching, which has led to the realization of spin~\cite{kronjager2010} or density~\cite{hung2013} wave patterns, or the study of supersolid order in BECs with spin orbit coupling ~\cite{Li17}, cavity mediated~\cite{Leonard17} or dipolar interactions~\cite{Bottcher19, Tanzi19, Chomaz18, Chomaz19, Zhang22}. 
It can also originate from parametric modulation, either of the trapping potential -- which can lead to the formation of waves~\cite{Engels07, Cominotti22} or vortices~\cite{Madison01, Sinha01} -- or the modulation of the interaction strength~\cite{Nguyen17, Nguyen19, Zhang20}.\\

The use of periodic parameter modulation to tailor the behavior of a quantum system lies at the heart of the broader and expanding field of Floquet engineering~\cite{Goldman14, Eckardt17, Weitenberg21}, where this modulation leads to effective Hamiltonians for the stroboscopic evolution. With ultracold atoms this approach has allowed to investigate a wide range of phenomena from effective dispersion relations~\cite{Lignier07, Kierig08, Ha15} to phase transitions~\cite{Parker13, Feng18, Song22}, and to the engineering of artificial magnetic fields and topological bands~\cite{Struck11,Aidelsburger11,Cooper19}.
In the context of Floquet engineering, parametric instabilities are a subject of particular interest, both as a potential source of heating and loss of coherence\cite{Lellouch17, Boulier19, Wintersperger20, Rubio20}, and as a way to direct pattern formation that may be used to reach specific quantum states~\cite{Campbell06, Galilo15, Engelhardt16, Zhang20}.

In this work, we exploit the tunability afforded by a Floquet system consisting of a Bose-Einstein condensate in a shaken 1D optical lattice, to control the appearance of a new, crystal-like order in the system through a parametric instability. The state produced is characterized by preserved coherence and a modulation of the density at a new spatial scale, spontaneously breaking the symmetry of the lattice, associated with the population of narrow, opposite, peaks in the momentum distribution. This is achieved through a lattice position modulation, resonant with interband transitions for initially empty modes. A Bogolubov analysis of the resulting effective system exhibits narrow instabilities in momentum space at opposite momenta, in the vicinity of avoided band crossings. This leads to the exponential growth, from fluctuations (which may be of quantum or thermal origin) of the population in modes with a narrow symmetrical momentum distribution. The position of these momentum components (and therefore the new long-range order in the system) is tunable through a change of the modulation parameters. The resulting state is only metastable, with further heating seemingly leading to its degradation. Our experimental observations are supported by simulations of the many body quantum dynamics with the truncated Wigner method, which exhibit the modulated density correlations and the preserved coherence of the resulting state. This is in contrast e.g. to the nucleation of staggered states~\cite{Gemelke05, Michon18, Mitchell21} where unstable modes have a fixed position and a broad momentum distribution, and where coherence is not preserved.

\begin{figure}
	\begin{center}
		\includegraphics[scale=0.9]{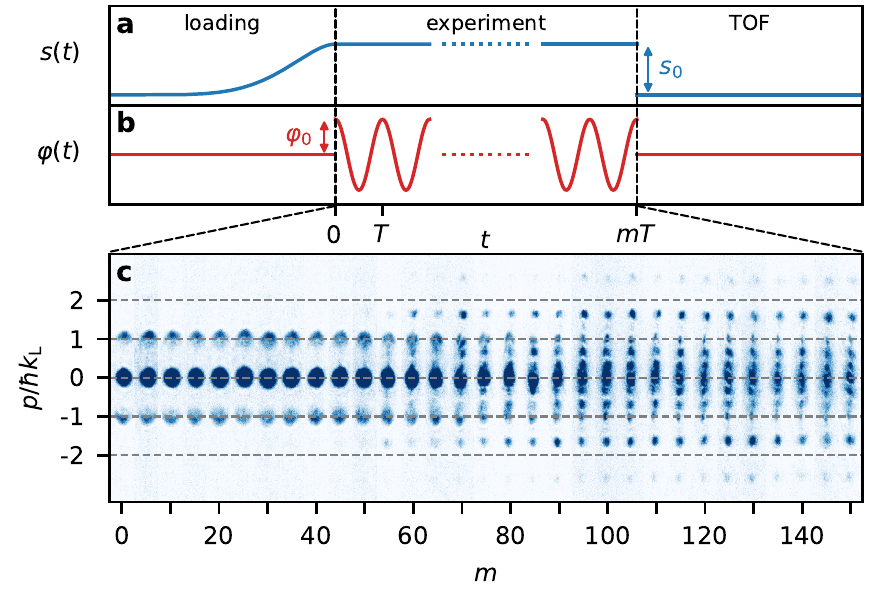}
		\caption{\textbf{Experimental protocol and typical results.} \textbf{(a-b)} Time-evolution of lattice depth (blue solid line) and phase (red solid line), showing the adiabatic loading to the ground state of the lattice, the phase-modulation experiment for an integer number of periods, and the lattice release and time-of-flight (TOF).
		\textbf{(c)} Stack of experimental absorption images after time-of-flight showing the stroboscopic evolution of the momentum distribution of an initial BEC with $N=10^5$ atoms, as a function of the number $m$ of periods of modulation $T$, with $s_0=3.4 \pm 0.10$, $\varphi_0=20 \degree$, $\nu = 1/T = 30$ kHz and $t_\dTOF= 35$ ms.}
		\label{fig:figure1}
	\end{center}
\end{figure}

\section*{Results}
\paragraph*{Experimental protocol.}
Our experimental study relies on a rubidium-87 Bose-Einstein condensate (BEC) machine that produces pure condensates with $5 \cdot 10^5$ atoms (unless otherwise stated) in a hybrid trap (see \cite{Fortun16} and Appendix~\ref{appen:setup} for more details). The BEC is adiabatically loaded in a far-detuned one-dimensional optical lattice of spacing $d=532$ nm resulting from the interference of two counter-propagating coherent laser beams along the $x-$axis. The potential experienced by the atoms reads

$$
V(x,y,z,t) = -\dfrac{s_0}{2}E_\dL \cos(k_\dL x + \varphi(t)) + U_\text{hyb}(x,y,z),
$$
where $k_\dL = 2\pi/d$, $s_0$ measures the lattice depth in units of $E_\dL = \hbar^2 k_\dL^2/2m$  and $U_\text{hyb}$ is the 3D harmonic potential of the hybrid trap characterized by the angular frequencies $(\omega_{x'}, \omega_{y'}, \omega_z) = 2\pi\times(10.4, 66, 68)$ Hz, where the axes $x',y'$ are at an $8\,\degree$ angle with $x,y$. 
The phase $\varphi(t)=\varphi_0\cos(2\pi \nu t)$ is modulated at a frequency resonant with interband transitions. The atoms also experience contact interactions characterized by the scattering length $a_\mathrm{s}$ ($a_\mathrm{s}\simeq5.3\,\mathrm{nm}$).

In Fig.\ref{fig:figure1}, the frequency $\nu=30$ kHz for $s_0\simeq 3.4$ couples the ground state band $s$ to the third excited band $f$ in the vicinity of quasi-momentum $q=\pm0.36k_\mathrm{L}$. The series of images shows the evolution of the matter wave diffraction pattern measured by absorption after a long time-of-flight ($t_\dTOF=35\,\mathrm{ms}$). It represents the stroboscopic evolution of the atomic momentum distribution in the modulated lattice at integer multiples $m$ of the modulation period. For short modulation times, we mostly observe three diffraction peaks centered on integer multiples of $h/d$ and associated to the initial ground state distribution of the static lattice. 
A faint halo is visible between the peaks,
which originates from elastic collisions occurring during the time-of-flight \cite{Greiner01,Tenart20,Chatelain_2020}. After 70 modulation periods ($\simeq 2.3\,\mathrm{ms}$), we clearly observe the emergence of symmetric diffraction peaks located in between the ordinary diffraction peaks. The initial growth of the population in the peaks appears to settle, with the new peaks remaining sharp over many modulation periods. Over longer timescales (see below), each narrow peak eventually seems to slowly broaden. These sharp peaks should be contrasted with the broader instabilities observed in the case of staggered states~\cite{Gemelke05, Michon18,Mitchell21} or single band parametric instabilities~\cite{Wintersperger20}.

\begin{figure}[t!]
	\begin{center}
		\includegraphics[scale=0.9]{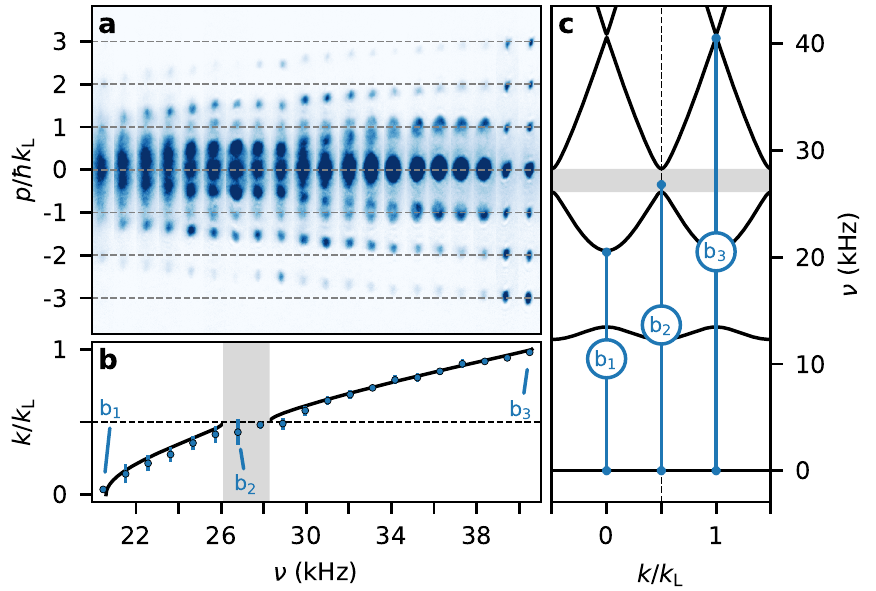}
		\caption{\textbf{Instability displacement with the modulation frequency.} \textbf{(a)} Stack of experimental absorption images after $n=100$ periods, averaged over 3 realizations, for an increasing modulation frequency $\nu$, with $s_0=3.57 \pm 0.10$, $\varphi_0= 15 \degree$ and $t_\dTOF= 35$ ms. \textbf{(b)} Average instability position (in terms of the reduced quasi-momentum $k/k_\mathrm{L}$ of the unfolded band structure) extracted from the fitted drift of the four $1 \leq |p|/\hbar k_\dL  \leq 3$ orders of diffraction over all realizations (blue dots, errorbars correspond to the standard deviation of the 12-point sample) and calculated position of the resonant coupling as a function of the modulation frequency (solid black line). \textbf{(c)} Transition diagram from the lowest band \textit{s} over the first two Brillouin zones (BZ) (solid black line) and addressed transitions for data b$_1$, b$_2$ and b$_3$ (blue dots and solid lines). In (b-c) the gap between the transitions \textit{s-d} and \textit{s-f} (grey shaded area) and edge of the first BZ  (black dotted line) are represented.}
		\label{fig:figure2}
	\end{center}
\end{figure}

\paragraph*{Tight-binding effective model and tunability.}
We interpret the emergent momentum peaks as originating from a parametric instability favoring a four-wave mixing process in which two atoms from the BEC with $q_\mathrm{in}=0$ scatter into quasi-momenta at $q=\pm q^*$ with $0<q^*<\pi/d$. This is associated with the emergence of a new spatial periodicity $d^*=\pi/q^*$ in the system. Correspondingly, the momentum distribution exhibits two families of peaks at momenta $p=\pm q^*+\ell k_L, \ell\in\mathbb{Z}$, related to the decomposition of the newly populated states in the band structure of the lattice potential.
This interpretation, and the physics at play, are correctly captured by an effective tight-binding model with two coupled bands, describing the Floquet system of lattice bands coupled by the lattice shaking (see Appendix~\ref{appen:model}). For realistic parameters, the Bogolubov treatment of this model reveals sharply localized unstable modes in the vicinity of the avoided band crossings. The position of the most unstable mode yields the central quasi-momenta $q=\pm q^*$.
From this modeling, we infer that the position of the instability in quasi-momentum smoothly depends on the model's parameters, namely the modulation frequency $\nu$ and amplitude $\varphi_0$, as well as the strength of interactions in the initial condensate.

In Fig.\ref{fig:figure2}, we compare the measured position (in terms of the reduced quasi-momentum $k/k_\mathrm{L}$ of the unfolded band structure) of the peaks emerging after a sufficiently long modulation time to the position of resonant transitions from the lowest band, as the frequency of modulation is varied. The modulation frequency is tuned across two bands, and is resonant with $s$ to $d$ transitions at low frequencies ($\nu < 26.1$ kHz), and with $s$ to $f$ transitions at higher frequencies ($\nu > 28.3$ kHz). As expected, the instability experimentally occurs in the vicinity of the band crossing. It seems however that the instability is systematically closer to the actual band crossing experimentally than expected from the model (which predicts it to be a few percent of $k_\mathrm{L}$ away from the crossing). Nevertheless, this demonstrates the tunability of the instability position in quasi-momentum space. The interpretation of the instability pattern for a transition frequency near the gap between bands $d$ and $f$ is more involved, as there is then a possibility of a resonant excitation of the condensate.

\paragraph*{Truncated Wigner simulation and correlations.}
To further investigate the coherence properties of the atomic state arising from the parametric instability, we perform numerical simulations of the modulated system, based on the Truncated Wigner method at zero temperature~\cite{TW1,TW2,TW3,TW4}. In these simulations the quantum state is expanded on the Wannier functions of the first five energy bands of the static lattice over a finite size system, taking into account the external confinement $U_\mathrm{hyb}$. This approach gives direct access to the first and second order correlation functions, revealing the features of the underlying many-body physics. After a sufficiently long evolution time, the spectrum of the one-body density matrix contains a dominant eigenvalue corresponding to the Bose Einstein condensate at zero quasi-momentum, plus two other significant eigenvalues corresponding to two states with opposite parity made of opposite quasi-momenta components (see Appendix~\ref{appen:TW}). 

\begin{figure}
	\begin{center}
		\includegraphics[scale=0.9]{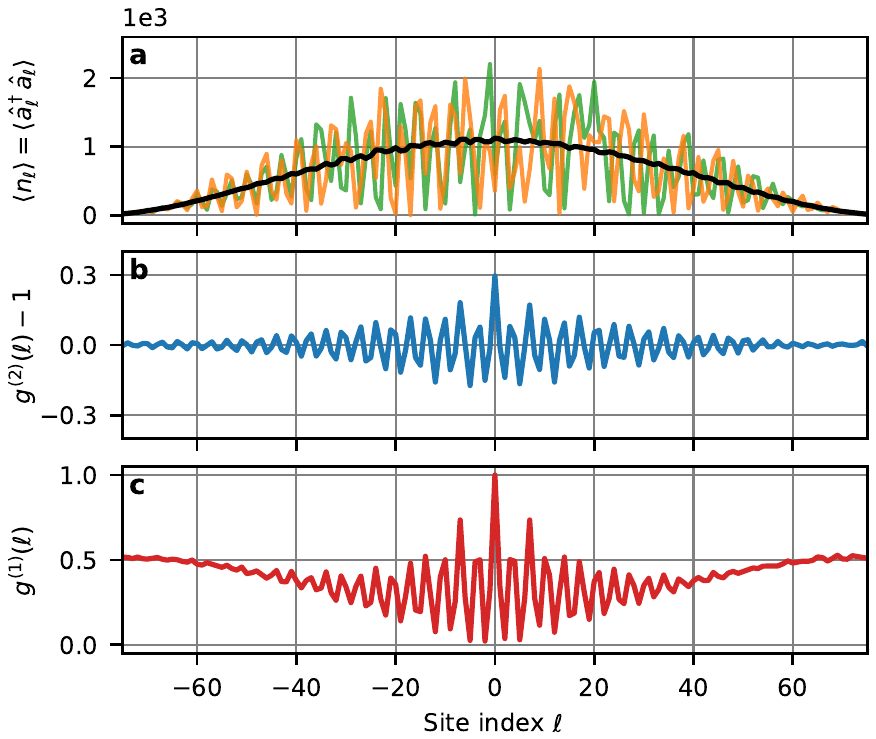}
		\caption{\textbf{Correlations and coherence after nucleation of the instability.} {\bf (a)} Density distribution in the $s$ band from TW simulations of a 1D modulated lattice with realistic trapping (see text for details). The chosen lattice parameters are $s_0=3.4$, $\nu=30\,\mathrm{kHz}$ and $\varphi_0=20^\circ$, with a number of atoms $N=10^5$. The density is  computed after a modulation duration of $5\,\mathrm{ms}$, when the instability is visible in the momentum distribution. Two individual TW trajectories (green and orange lines) exhibit a clear periodic modulation of the density, which is washed out in the average of $1,000$ trajectories (thick black line). {\bf (b-c)} In the same conditions as {\bf (a)}, average density correlation function $g^{(2)}(\ell)$  {\bf (b)} and average amplitude correlation function $g^{(1)}(\ell)$ {\bf (c)} between the central site and a site $\ell$ in the $s$ band of the lattice potential (see text for definitions).}
		\label{fig:figure4}
	\end{center}
\end{figure}

The macroscopic occupation of modes having opposite quasi-momenta in the Brillouin zone raises the question of the possible emergence of a long-range order in the system. To answer this question, we exploit our beyond-mean-field calculations to compute the atom density on each site across the system. Strikingly, in Fig.\ref{fig:figure4}{\bf(a)}, after $5\,$ms of evolution, we see the evidence on individual TW trajectories of the emergence of a new periodic modulation of the density within the lowest band of the lattice, on a scale of about 3 lattice sites. This is related to the quasi-momentum at which the instability occurs, which is $q^*\simeq 0.29$, leading to a spacing $d^*/d\simeq 3.4$, which is as mentioned before is very close to the experimental finding of $q*=\pm0.36$ for the same set of parameters. In each trajectory from the TW simulation, the phase reference of the density modulation appears random, which washes out the modulation on the average density (full line in Fig.\ref{fig:figure4}{\bf(a)}). We can recover evidence for the modulation by computing the normalized density-density correlation function in the lowest band $g^{(2)}(\ell)=\langle n_0 n_\ell\rangle / ( \langle n_0\rangle \langle n_\ell\rangle)$, between the central site of the lattice and a site $\ell$, averaged over $1,000$ TW trajectories. This is plotted in Fig.\ref{fig:figure4}{\bf(b)} as a function of the site position $\ell$. It reveals strong, regular, density oscillations across the whole system which signal the emergence of a new long-range order. Finally in Fig.\ref{fig:figure4}{\bf(c)} we similarly compute a normalized coherence within the ground band between sites $0$ and $\ell$ in the system $g^{(1)}(\ell)=\langle \hat{a}_0^\dagger\hat{a}_\ell\rangle / (\langle n_0\rangle \langle n_\ell\rangle)^{1/2}$. This shows modulations with a similar period, demonstrating the coexistence across the system of the initial condensate of period $d$, as well as fully-coherent states with the new period $d^*$ (Appendix~\ref{appen:TW}). This coherence is further demonstrated by the time-of-flight diffraction pattern. 

The appearance of these modulations is actually already present within the tight-binding model, in an estimate of the correlation functions (see Appendix~\ref{appen:model}). This estimate allows us to highlight the fact that the appearance of this new spatial order is intimately related to the sharpness of the instability maximum in momentum space, which distinguishes the resulting state from staggered states~\cite{Gemelke05, Michon18,Mitchell21}. These properties of the correlation functions are reminiscent of those of a supersolid state, however it must be emphasized that the produced state is metastable and cannot to our knowledge  be characterized as the ground state of an effective Hamiltonian, thereby lacking a key property to be qualified a supersolid.

\begin{figure}
	\begin{center}
		\includegraphics[scale=0.9]{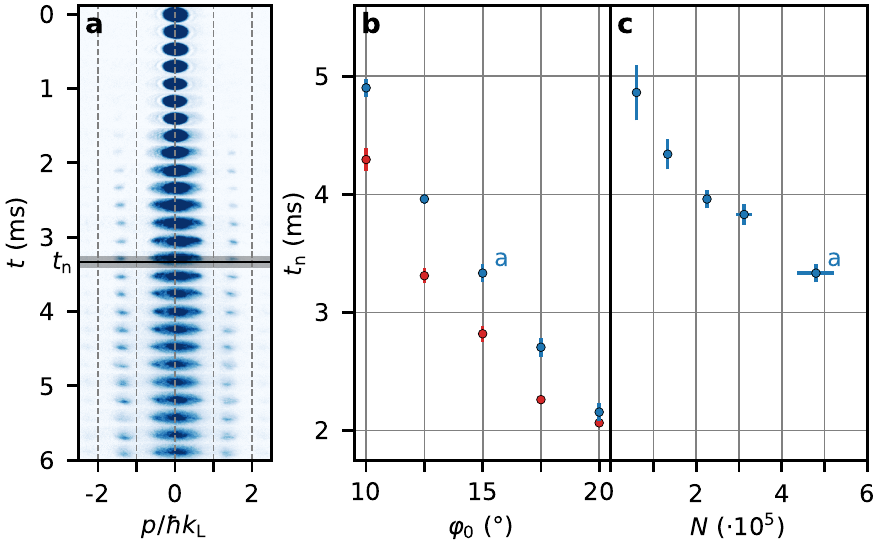}
		\caption{\textbf{Nucleation time of the instability.} \textbf{(a)} Experimental record for determining the nucleation timescale : using band-mapping before imaging leads to well-isolated contributions from higher lattice bands (here for $1<|p|/(\hbar k_L)<2$) signaling the growth of the instability. This growth is fitted to extract a nucleation time $t_\mathrm{n}$, denoted by the horizontal line, with uncertainty represented by the shaded area (see text for details). \textbf{(b)} Evolution of the measured nucleation time of the unstable modes (see text) as a function of the modulation amplitude $\varphi_0$ for $\nu=25.5$ kHz and $s_0=3.70 \pm 0.20$ (blue, coupling bands \textit{s-d}) and $\nu=30.5$ kHz and $s_0=3.56 \pm 0.20$ (red, coupling bands \textit{s-f}). Error bars correspond to one standard deviation on the fit (see text). \textbf{(c)} Same as \textbf{(b)} as a function of the number of atoms $N$ in the BEC for $\nu=25.5$ kHz and $s_0=3.58\pm0.30$, for a fixed modulation amplitude $ \varphi_0=15\degree$. Error bars correspond vertically to one standard deviation on the fit and horizontally to one standard deviation over 4 independent measurements of the atom number.}
		\label{fig:figure3}
	\end{center}
\end{figure}

\paragraph*{Nucleation timescale.}
The two-band model can provide an estimate for the value of the instability exponents, and therefore the timescale, that characterize the exponential growth of unstable modes, as well as for their position. But it cannot provide the full dynamics of the mode growth nor their subsequent evolution. However, the real timescale information can be readily accessed experimentally. In Fig.\ref{fig:figure3}, we measured the nucleation time $t_\mathrm{n}$ as a function of the modulation amplitude $\varphi_0$ and the number of atoms $N$ in the BEC. To precisely extract the value of $t_\mathrm{n}$, we used a band-mapping technique (see Appendix~\ref{appen:bandmap}): after modulating the lattice for an integer number of modulation periods, the modulation is stopped, and the lattice depth adiabatically lowered before performing a time-of-flight. This technique allows us to unambiguously identify the population of modes in the excited bands at finite quasi-momenta resulting from the instability, and reduces the impact of remnant collision halos on the measurement compared with a direct time-of-flight measurement. The growth of the population in the higher band modes is fitted with a sigmoid growth curve to yield a nucleation time at the half-maximum point.

Our results in Fig.\ref{fig:figure3} show that the nucleation time decreases with the modulation amplitude and the number of atoms $N$ (experimentally, the number of atoms in the BEC can be reduced by a factor of up to ten in a reproducible manner by evaporating the BEC held in the hybrid trap before loading the lattice). These trends are similar to those observed for the single band Bogolubov instability leading to staggered states~\cite{Michon18}. They also are expected from trends in the tight-binding model for realistic parameters (see Appendix~\ref{appen:model}): by increasing the modulation amplitude $\varphi_0$, we increase the coupling between bands, which leads to larger instability exponents. The variation of the nucleation time with $N$ is qualitatively expected as well, as the initial interaction energy in the condensate increases with $N$, and is also associated with a stronger instability. The TW simulations provide a more accurate and complete description of the evolution of the system, that should be qualitatively correct (within the approximations of the model). That is indeed the case: with the shared parameters of Fig.\ref{fig:figure1} (experiment) and Fig.\ref{fig:figure4} (simulation), the instability occurs at the same position in momentum space. The nucleation time in simulations is longer than in experiments, by about a factor of 2 : this may be due to a thermal activation in the experiment (the simulations are at zero-temperature), or a contribution of the transverse degrees of freedom, that are not included in the 1D simulations.

We restricted our measurements to small values of $\varphi_0$, beyond which the validity of the two-band model and of our interpretation becomes questionable. Indeed, for stronger coupling values, the Floquet spectrum increasingly contains significant avoided crossings and fully hybridized bands, and the evolution of the system is expected to get more complex. 

\section*{Discussion}

\begin{figure}
	\begin{center}
		\includegraphics[scale=0.9]{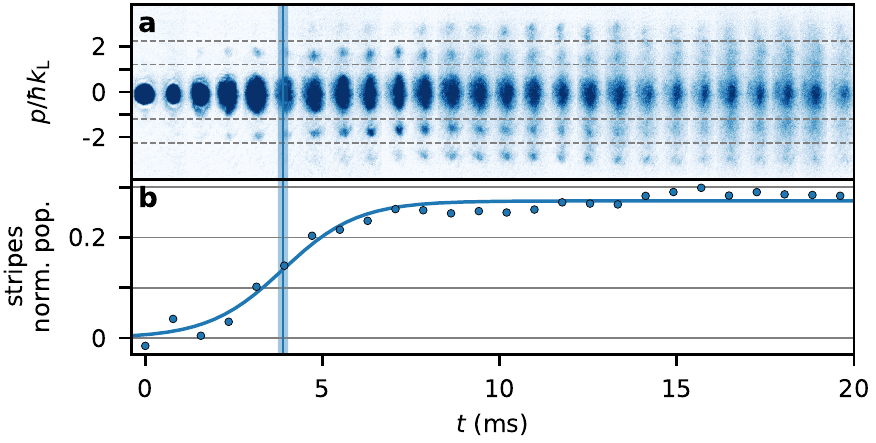}
		\caption{\textbf{Long term survival of the emerging crystalline order} \textbf{(a)} Experimental stroboscopic evolution of the momentum distribution after bandmapping over 510 periods of modulation. Modulation parameters are $s_0=3.8\pm0.1$, $\nu=25.5\,\mathrm{kHz}$ and $\varphi_0=15\,\degree$. The initial growth of the instability is monitored by counting the relative atomic population in the two stripes delimited by dashed lines. \textbf{(b)} Evolution of the normalized population in momentum peaks created by the instability (stripes in \textbf{(a)}). A sigmoid fit allows to determine the nucleation timescale $t_\mathrm{n}\simeq4\,\mathrm{ms}$. The population in the nucleated modes then settles, and remains identifiable on a timescale of several $t_\mathrm{n}$, but an eventual broadening of the momentum peaks is clearly visible in \textbf{(a)}.}
		\label{fig:survival}
	\end{center}
\end{figure}

We have also investigated the survival of the produced state after saturation of the instability. This is illustrated in Figure~\ref{fig:survival}. As mentioned earlier, after the population of the newly formed peaks in the momentum distribution settles (as seen from the bandmapping data), the peaks at $q=\pm q^*$ seem to slowly broaden, presumably through some further instability that leads to heating. Interestingly this heating effect seems more pronounced in simulations than in experiments. In simulations, the sharp peaks at $q=\pm q^*$ only exist in a small time interval around $t_\mathrm{n}$, while experimentally (Fig.~\ref{fig:survival}) we can reliably see the contributions at $\pm q^*$ after nucleation over a duration of order $~3t_\mathrm{n}$. We hypothesize that this longer persistence in experiments than in numerical simulations may originate from the transverse degrees of freedom in the experimental system (absent in the simulation), which offer more possibilities for effective heat dissipation, and can affect parametric instabilities~\cite{Wintersperger20}.

In summary, we have investigated how a tunable crystalline order emerges from atom-atom interactions seeded by fluctuations in a Floquet system generated by resonant band coupling.  The tunable long-range order observed here is synthesized and controlled by the parameters of the modulation and an atomic four-wave mixing effect stemming from contact interactions in the ultracold regime. It is as such applicable to other atomic species or systems with contact interactions.
A key feature of this instability is the ability, via the coupling to higher bands, to tailor both the position and width of the instability, which adjusts the periodicity and range of first and second order correlations.
Further investigations into this effect can be envisioned, to characterize the ulterior broadening of the momentum components and the stability of the system, to possibly observe in situ the density modulation, e.g. through the quantum gas magnifier technique~\cite{Asteria21,Zahn22}, or to explore similar effects in higher dimensions and/or other lattice geometries. This work highlights the potential of the interplay between Floquet engineering and interactions for the preparation and manipulation of exotic quantum states.

\medskip

\paragraph*{Acknowledgments.}

The authors thank Jean Dalibard for helpful discussions. This work was supported by research funding Grant No. ANR-17-CE30-0024. ND and FA acknowledge support from Région Occitanie and Université Toulouse III-Paul Sabatier. MA acknowledges support from the DGA (Direction G\'en\'erale de l’Armement).

\paragraph*{Authors contributions.}

DGO, BP and JB designed the research and supervised the experimental work. ND, LG and FA  performed the experiments presented in the manuscript, the data collection and analysis. GC and MA obtained the first experimental observations of the instability. PS conceived, developed and performed the TW simulations. BP and PS developed the tight-binding model analysis. DGO and BP coordinated the exchanges between experiment and numerical simulations. DGO, BP and PS  wrote the manuscript. All authors discussed and contributed to the final version of the manuscript. 


\paragraph*{Competing interests.}

The authors declare no competing interests.

\paragraph*{Data availability.}

The data that support the findings of this study are available from the corresponding author upon reasonable request.

\bibliography{article_hybrid_patternsNotes}

\begin{thebibliography}{57}%
\makeatletter
\providecommand \@ifxundefined [1]{%
 \@ifx{#1\undefined}
}%
\providecommand \@ifnum [1]{%
 \ifnum #1\expandafter \@firstoftwo
 \else \expandafter \@secondoftwo
 \fi
}%
\providecommand \@ifx [1]{%
 \ifx #1\expandafter \@firstoftwo
 \else \expandafter \@secondoftwo
 \fi
}%
\providecommand \natexlab [1]{#1}%
\providecommand \enquote  [1]{``#1''}%
\providecommand \bibnamefont  [1]{#1}%
\providecommand \bibfnamefont [1]{#1}%
\providecommand \citenamefont [1]{#1}%
\providecommand \href@noop [0]{\@secondoftwo}%
\providecommand \href [0]{\begingroup \@sanitize@url \@href}%
\providecommand \@href[1]{\@@startlink{#1}\@@href}%
\providecommand \@@href[1]{\endgroup#1\@@endlink}%
\providecommand \@sanitize@url [0]{\catcode `\\12\catcode `\$12\catcode
  `\&12\catcode `\#12\catcode `\^12\catcode `\_12\catcode `\%12\relax}%
\providecommand \@@startlink[1]{}%
\providecommand \@@endlink[0]{}%
\providecommand \url  [0]{\begingroup\@sanitize@url \@url }%
\providecommand \@url [1]{\endgroup\@href {#1}{\urlprefix }}%
\providecommand \urlprefix  [0]{URL }%
\providecommand \Eprint [0]{\href }%
\providecommand \doibase [0]{https://doi.org/}%
\providecommand \selectlanguage [0]{\@gobble}%
\providecommand \bibinfo  [0]{\@secondoftwo}%
\providecommand \bibfield  [0]{\@secondoftwo}%
\providecommand \translation [1]{[#1]}%
\providecommand \BibitemOpen [0]{}%
\providecommand \bibitemStop [0]{}%
\providecommand \bibitemNoStop [0]{.\EOS\space}%
\providecommand \EOS [0]{\spacefactor3000\relax}%
\providecommand \BibitemShut  [1]{\csname bibitem#1\endcsname}%
\let\auto@bib@innerbib\@empty
\bibitem [{\citenamefont {Lopes}\ \emph {et~al.}(2017)\citenamefont {Lopes},
  \citenamefont {Eigen}, \citenamefont {Navon}, \citenamefont {Cl\'ement},
  \citenamefont {Smith},\ and\ \citenamefont {Hadzibabic}}]{Lopes17}%
  \BibitemOpen
  \bibfield  {author} {\bibinfo {author} {\bibfnamefont {R.}~\bibnamefont
  {Lopes}}, \bibinfo {author} {\bibfnamefont {C.}~\bibnamefont {Eigen}},
  \bibinfo {author} {\bibfnamefont {N.}~\bibnamefont {Navon}}, \bibinfo
  {author} {\bibfnamefont {D.}~\bibnamefont {Cl\'ement}}, \bibinfo {author}
  {\bibfnamefont {R.~P.}\ \bibnamefont {Smith}},\ and\ \bibinfo {author}
  {\bibfnamefont {Z.}~\bibnamefont {Hadzibabic}},\ }\bibfield  {title}
  {\bibinfo {title} {{Quantum Depletion of a Homogeneous Bose-Einstein
  Condensate}},\ }\href {https://doi.org/10.1103/PhysRevLett.119.190404}
  {\bibfield  {journal} {\bibinfo  {journal} {Phys. Rev. Lett.}\ }\textbf
  {\bibinfo {volume} {119}},\ \bibinfo {pages} {190404} (\bibinfo {year}
  {2017})}\BibitemShut {NoStop}%
\bibitem [{\citenamefont {Xu}\ \emph {et~al.}(2006)\citenamefont {Xu},
  \citenamefont {Liu}, \citenamefont {Miller}, \citenamefont {Chin},
  \citenamefont {Setiawan},\ and\ \citenamefont {Ketterle}}]{Xu06}%
  \BibitemOpen
  \bibfield  {author} {\bibinfo {author} {\bibfnamefont {K.}~\bibnamefont
  {Xu}}, \bibinfo {author} {\bibfnamefont {Y.}~\bibnamefont {Liu}}, \bibinfo
  {author} {\bibfnamefont {D.~E.}\ \bibnamefont {Miller}}, \bibinfo {author}
  {\bibfnamefont {J.~K.}\ \bibnamefont {Chin}}, \bibinfo {author}
  {\bibfnamefont {W.}~\bibnamefont {Setiawan}},\ and\ \bibinfo {author}
  {\bibfnamefont {W.}~\bibnamefont {Ketterle}},\ }\bibfield  {title} {\bibinfo
  {title} {{Observation of Strong Quantum Depletion in a Gaseous Bose-Einstein
  Condensate}},\ }\href {https://doi.org/10.1103/PhysRevLett.96.180405}
  {\bibfield  {journal} {\bibinfo  {journal} {Phys. Rev. Lett.}\ }\textbf
  {\bibinfo {volume} {96}},\ \bibinfo {pages} {180405} (\bibinfo {year}
  {2006})}\BibitemShut {NoStop}%
\bibitem [{\citenamefont {Cayla}\ \emph {et~al.}(2020)\citenamefont {Cayla},
  \citenamefont {Butera}, \citenamefont {Carcy}, \citenamefont {Tenart},
  \citenamefont {Herc\'e}, \citenamefont {Mancini}, \citenamefont {Aspect},
  \citenamefont {Carusotto},\ and\ \citenamefont {Cl\'ement}}]{Cayla20}%
  \BibitemOpen
  \bibfield  {author} {\bibinfo {author} {\bibfnamefont {H.}~\bibnamefont
  {Cayla}}, \bibinfo {author} {\bibfnamefont {S.}~\bibnamefont {Butera}},
  \bibinfo {author} {\bibfnamefont {C.}~\bibnamefont {Carcy}}, \bibinfo
  {author} {\bibfnamefont {A.}~\bibnamefont {Tenart}}, \bibinfo {author}
  {\bibfnamefont {G.}~\bibnamefont {Herc\'e}}, \bibinfo {author} {\bibfnamefont
  {M.}~\bibnamefont {Mancini}}, \bibinfo {author} {\bibfnamefont
  {A.}~\bibnamefont {Aspect}}, \bibinfo {author} {\bibfnamefont
  {I.}~\bibnamefont {Carusotto}},\ and\ \bibinfo {author} {\bibfnamefont
  {D.}~\bibnamefont {Cl\'ement}},\ }\bibfield  {title} {\bibinfo {title}
  {{Hanbury Brown and Twiss Bunching of Phonons and of the Quantum Depletion in
  an Interacting Bose Gas}},\ }\href
  {https://doi.org/10.1103/PhysRevLett.125.165301} {\bibfield  {journal}
  {\bibinfo  {journal} {Phys. Rev. Lett.}\ }\textbf {\bibinfo {volume} {125}},\
  \bibinfo {pages} {165301} (\bibinfo {year} {2020})}\BibitemShut {NoStop}%
\bibitem [{\citenamefont {Tenart}\ \emph {et~al.}(2021)\citenamefont {Tenart},
  \citenamefont {Herc{\'e}}, \citenamefont {Bureik}, \citenamefont {Dareau},\
  and\ \citenamefont {Cl{\'e}ment}}]{Tenart21}%
  \BibitemOpen
  \bibfield  {author} {\bibinfo {author} {\bibfnamefont {A.}~\bibnamefont
  {Tenart}}, \bibinfo {author} {\bibfnamefont {G.}~\bibnamefont {Herc{\'e}}},
  \bibinfo {author} {\bibfnamefont {J.-P.}\ \bibnamefont {Bureik}}, \bibinfo
  {author} {\bibfnamefont {A.}~\bibnamefont {Dareau}},\ and\ \bibinfo {author}
  {\bibfnamefont {D.}~\bibnamefont {Cl{\'e}ment}},\ }\bibfield  {title}
  {\bibinfo {title} {Observation of pairs of atoms at opposite momenta in an
  equilibrium interacting {Bose} gas},\ }\href
  {https://doi.org/10.1038/s41567-021-01381-2} {\bibfield  {journal} {\bibinfo
  {journal} {Nature Physics}\ }\textbf {\bibinfo {volume} {17}},\ \bibinfo
  {pages} {1364} (\bibinfo {year} {2021})}\BibitemShut {NoStop}%
\bibitem [{\citenamefont {Kronjäger}\ \emph {et~al.}(2010)\citenamefont
  {Kronjäger}, \citenamefont {Becker}, \citenamefont {Soltan-Panahi},
  \citenamefont {Bongs},\ and\ \citenamefont {Sengstock}}]{kronjager2010}%
  \BibitemOpen
  \bibfield  {author} {\bibinfo {author} {\bibfnamefont {J.}~\bibnamefont
  {Kronjäger}}, \bibinfo {author} {\bibfnamefont {C.}~\bibnamefont {Becker}},
  \bibinfo {author} {\bibfnamefont {P.}~\bibnamefont {Soltan-Panahi}}, \bibinfo
  {author} {\bibfnamefont {K.}~\bibnamefont {Bongs}},\ and\ \bibinfo {author}
  {\bibfnamefont {K.}~\bibnamefont {Sengstock}},\ }\bibfield  {title} {\bibinfo
  {title} {Spontaneous {Pattern} {Formation} in an {Antiferromagnetic}
  {Quantum} {Gas}},\ }\href {https://doi.org/10.1103/PhysRevLett.105.090402}
  {\bibfield  {journal} {\bibinfo  {journal} {Physical Review Letters}\
  }\textbf {\bibinfo {volume} {105}},\ \bibinfo {pages} {090402} (\bibinfo
  {year} {2010})}\BibitemShut {NoStop}%
\bibitem [{\citenamefont {Hung}\ \emph {et~al.}(2013)\citenamefont {Hung},
  \citenamefont {Gurarie},\ and\ \citenamefont {Chin}}]{hung2013}%
  \BibitemOpen
  \bibfield  {author} {\bibinfo {author} {\bibfnamefont {C.-L.}\ \bibnamefont
  {Hung}}, \bibinfo {author} {\bibfnamefont {V.}~\bibnamefont {Gurarie}},\ and\
  \bibinfo {author} {\bibfnamefont {C.}~\bibnamefont {Chin}},\ }\bibfield
  {title} {\bibinfo {title} {From {Cosmology} to {Cold} {Atoms}: {Observation}
  of {Sakharov} {Oscillations} in a {Quenched} {Atomic} {Superfluid}},\ }\href
  {https://doi.org/10.1126/science.1237557} {\bibfield  {journal} {\bibinfo
  {journal} {Science}\ }\textbf {\bibinfo {volume} {341}},\ \bibinfo {pages}
  {1213} (\bibinfo {year} {2013})}\BibitemShut {NoStop}%
\bibitem [{\citenamefont {Li}\ \emph {et~al.}(2017)\citenamefont {Li},
  \citenamefont {Lee}, \citenamefont {Huang}, \citenamefont {Burchesky},
  \citenamefont {Shteynas}, \citenamefont {Top}, \citenamefont {Jamison},\ and\
  \citenamefont {Ketterle}}]{Li17}%
  \BibitemOpen
  \bibfield  {author} {\bibinfo {author} {\bibfnamefont {J.-R.}\ \bibnamefont
  {Li}}, \bibinfo {author} {\bibfnamefont {J.}~\bibnamefont {Lee}}, \bibinfo
  {author} {\bibfnamefont {W.}~\bibnamefont {Huang}}, \bibinfo {author}
  {\bibfnamefont {S.}~\bibnamefont {Burchesky}}, \bibinfo {author}
  {\bibfnamefont {B.}~\bibnamefont {Shteynas}}, \bibinfo {author}
  {\bibfnamefont {F.-C.}\ \bibnamefont {Top}}, \bibinfo {author} {\bibfnamefont
  {A.}~\bibnamefont {Jamison}},\ and\ \bibinfo {author} {\bibfnamefont
  {W.}~\bibnamefont {Ketterle}},\ }\bibfield  {title} {\bibinfo {title} {A
  stripe phase with supersolid properties in spin–orbit-coupled{
  Bose–Einstein} condensates},\ }\href
  {https://doi.org/https://doi.org/10.1038/nature21431} {\bibfield  {journal}
  {\bibinfo  {journal} {Nature}\ }\textbf {\bibinfo {volume} {543}},\ \bibinfo
  {pages} {91} (\bibinfo {year} {2017})}\BibitemShut {NoStop}%
\bibitem [{\citenamefont {L{\'e}onard}\ \emph {et~al.}(2017)\citenamefont
  {L{\'e}onard}, \citenamefont {Morales}, \citenamefont {Zupancic},
  \citenamefont {Esslinger},\ and\ \citenamefont {Donner}}]{Leonard17}%
  \BibitemOpen
  \bibfield  {author} {\bibinfo {author} {\bibfnamefont {J.}~\bibnamefont
  {L{\'e}onard}}, \bibinfo {author} {\bibfnamefont {A.}~\bibnamefont
  {Morales}}, \bibinfo {author} {\bibfnamefont {P.}~\bibnamefont {Zupancic}},
  \bibinfo {author} {\bibfnamefont {T.}~\bibnamefont {Esslinger}},\ and\
  \bibinfo {author} {\bibfnamefont {T.}~\bibnamefont {Donner}},\ }\bibfield
  {title} {\bibinfo {title} {Supersolid formation in a quantum gas breaking a
  continuous translational symmetry},\ }\href
  {https://doi.org/https://doi.org/10.1038/nature21067} {\bibfield  {journal}
  {\bibinfo  {journal} {Nature}\ }\textbf {\bibinfo {volume} {543}},\ \bibinfo
  {pages} {87} (\bibinfo {year} {2017})}\BibitemShut {NoStop}%
\bibitem [{\citenamefont {B\"ottcher}\ \emph {et~al.}(2019)\citenamefont
  {B\"ottcher}, \citenamefont {Schmidt}, \citenamefont {Wenzel}, \citenamefont
  {Hertkorn}, \citenamefont {Guo}, \citenamefont {Langen},\ and\ \citenamefont
  {Pfau}}]{Bottcher19}%
  \BibitemOpen
  \bibfield  {author} {\bibinfo {author} {\bibfnamefont {F.}~\bibnamefont
  {B\"ottcher}}, \bibinfo {author} {\bibfnamefont {J.-N.}\ \bibnamefont
  {Schmidt}}, \bibinfo {author} {\bibfnamefont {M.}~\bibnamefont {Wenzel}},
  \bibinfo {author} {\bibfnamefont {J.}~\bibnamefont {Hertkorn}}, \bibinfo
  {author} {\bibfnamefont {M.}~\bibnamefont {Guo}}, \bibinfo {author}
  {\bibfnamefont {T.}~\bibnamefont {Langen}},\ and\ \bibinfo {author}
  {\bibfnamefont {T.}~\bibnamefont {Pfau}},\ }\bibfield  {title} {\bibinfo
  {title} {{Transient Supersolid Properties in an Array of Dipolar Quantum
  Droplets}},\ }\href {https://doi.org/10.1103/PhysRevX.9.011051} {\bibfield
  {journal} {\bibinfo  {journal} {Phys. Rev. X}\ }\textbf {\bibinfo {volume}
  {9}},\ \bibinfo {pages} {011051} (\bibinfo {year} {2019})}\BibitemShut
  {NoStop}%
\bibitem [{\citenamefont {Tanzi}\ \emph {et~al.}(2019)\citenamefont {Tanzi},
  \citenamefont {Lucioni}, \citenamefont {Fam\`a}, \citenamefont {Catani},
  \citenamefont {Fioretti}, \citenamefont {Gabbanini}, \citenamefont {Bisset},
  \citenamefont {Santos},\ and\ \citenamefont {Modugno}}]{Tanzi19}%
  \BibitemOpen
  \bibfield  {author} {\bibinfo {author} {\bibfnamefont {L.}~\bibnamefont
  {Tanzi}}, \bibinfo {author} {\bibfnamefont {E.}~\bibnamefont {Lucioni}},
  \bibinfo {author} {\bibfnamefont {F.}~\bibnamefont {Fam\`a}}, \bibinfo
  {author} {\bibfnamefont {J.}~\bibnamefont {Catani}}, \bibinfo {author}
  {\bibfnamefont {A.}~\bibnamefont {Fioretti}}, \bibinfo {author}
  {\bibfnamefont {C.}~\bibnamefont {Gabbanini}}, \bibinfo {author}
  {\bibfnamefont {R.~N.}\ \bibnamefont {Bisset}}, \bibinfo {author}
  {\bibfnamefont {L.}~\bibnamefont {Santos}},\ and\ \bibinfo {author}
  {\bibfnamefont {G.}~\bibnamefont {Modugno}},\ }\bibfield  {title} {\bibinfo
  {title} {{Observation of a Dipolar Quantum Gas with Metastable Supersolid
  Properties}},\ }\href {https://doi.org/10.1103/PhysRevLett.122.130405}
  {\bibfield  {journal} {\bibinfo  {journal} {Phys. Rev. Lett.}\ }\textbf
  {\bibinfo {volume} {122}},\ \bibinfo {pages} {130405} (\bibinfo {year}
  {2019})}\BibitemShut {NoStop}%
\bibitem [{\citenamefont {Chomaz}\ \emph {et~al.}(2018)\citenamefont {Chomaz},
  \citenamefont {van Bijnen}, \citenamefont {Petter}, \citenamefont {Faraoni},
  \citenamefont {Baier}, \citenamefont {Becher}, \citenamefont {Mark},
  \citenamefont {Wächtler}, \citenamefont {Santos},\ and\ \citenamefont
  {Ferlaino}}]{Chomaz18}%
  \BibitemOpen
  \bibfield  {author} {\bibinfo {author} {\bibfnamefont {L.}~\bibnamefont
  {Chomaz}}, \bibinfo {author} {\bibfnamefont {R.~M.~W.}\ \bibnamefont {van
  Bijnen}}, \bibinfo {author} {\bibfnamefont {D.}~\bibnamefont {Petter}},
  \bibinfo {author} {\bibfnamefont {G.}~\bibnamefont {Faraoni}}, \bibinfo
  {author} {\bibfnamefont {S.}~\bibnamefont {Baier}}, \bibinfo {author}
  {\bibfnamefont {J.~H.}\ \bibnamefont {Becher}}, \bibinfo {author}
  {\bibfnamefont {M.~J.}\ \bibnamefont {Mark}}, \bibinfo {author}
  {\bibfnamefont {F.}~\bibnamefont {Wächtler}}, \bibinfo {author}
  {\bibfnamefont {L.}~\bibnamefont {Santos}},\ and\ \bibinfo {author}
  {\bibfnamefont {F.}~\bibnamefont {Ferlaino}},\ }\bibfield  {title} {\bibinfo
  {title} {Observation of roton mode population in a dipolar quantum gas},\
  }\href {https://doi.org/10.1038/s41567-018-0054-7} {\bibfield  {journal}
  {\bibinfo  {journal} {Nature Physics}\ }\textbf {\bibinfo {volume} {14}},\
  \bibinfo {pages} {442} (\bibinfo {year} {2018})}\BibitemShut {NoStop}%
\bibitem [{\citenamefont {Chomaz}\ \emph {et~al.}(2019)\citenamefont {Chomaz},
  \citenamefont {Petter}, \citenamefont {Ilzh\"ofer}, \citenamefont {Natale},
  \citenamefont {Trautmann}, \citenamefont {Politi}, \citenamefont
  {Durastante}, \citenamefont {van Bijnen}, \citenamefont {Patscheider},
  \citenamefont {Sohmen}, \citenamefont {Mark},\ and\ \citenamefont
  {Ferlaino}}]{Chomaz19}%
  \BibitemOpen
  \bibfield  {author} {\bibinfo {author} {\bibfnamefont {L.}~\bibnamefont
  {Chomaz}}, \bibinfo {author} {\bibfnamefont {D.}~\bibnamefont {Petter}},
  \bibinfo {author} {\bibfnamefont {P.}~\bibnamefont {Ilzh\"ofer}}, \bibinfo
  {author} {\bibfnamefont {G.}~\bibnamefont {Natale}}, \bibinfo {author}
  {\bibfnamefont {A.}~\bibnamefont {Trautmann}}, \bibinfo {author}
  {\bibfnamefont {C.}~\bibnamefont {Politi}}, \bibinfo {author} {\bibfnamefont
  {G.}~\bibnamefont {Durastante}}, \bibinfo {author} {\bibfnamefont {R.~M.~W.}\
  \bibnamefont {van Bijnen}}, \bibinfo {author} {\bibfnamefont
  {A.}~\bibnamefont {Patscheider}}, \bibinfo {author} {\bibfnamefont
  {M.}~\bibnamefont {Sohmen}}, \bibinfo {author} {\bibfnamefont {M.~J.}\
  \bibnamefont {Mark}},\ and\ \bibinfo {author} {\bibfnamefont
  {F.}~\bibnamefont {Ferlaino}},\ }\bibfield  {title} {\bibinfo {title}
  {{Long-Lived and Transient Supersolid Behaviors in Dipolar Quantum Gases}},\
  }\href {https://doi.org/10.1103/PhysRevX.9.021012} {\bibfield  {journal}
  {\bibinfo  {journal} {Phys. Rev. X}\ }\textbf {\bibinfo {volume} {9}},\
  \bibinfo {pages} {021012} (\bibinfo {year} {2019})}\BibitemShut {NoStop}%
\bibitem [{\citenamefont {Zhang}\ \emph {et~al.}(2022)\citenamefont {Zhang},
  \citenamefont {Zhang}, \citenamefont {Yang},\ and\ \citenamefont
  {Capogrosso-Sansone}}]{Zhang22}%
  \BibitemOpen
  \bibfield  {author} {\bibinfo {author} {\bibfnamefont {J.}~\bibnamefont
  {Zhang}}, \bibinfo {author} {\bibfnamefont {C.}~\bibnamefont {Zhang}},
  \bibinfo {author} {\bibfnamefont {J.}~\bibnamefont {Yang}},\ and\ \bibinfo
  {author} {\bibfnamefont {B.}~\bibnamefont {Capogrosso-Sansone}},\ }\bibfield
  {title} {\bibinfo {title} {Supersolid phases of lattice dipoles tilted in
  three dimensions},\ }\href {https://doi.org/10.1103/PhysRevA.105.063302}
  {\bibfield  {journal} {\bibinfo  {journal} {Physical Review A}\ }\textbf
  {\bibinfo {volume} {105}},\ \bibinfo {pages} {063302} (\bibinfo {year}
  {2022})}\BibitemShut {NoStop}%
\bibitem [{\citenamefont {Engels}\ \emph {et~al.}(2007)\citenamefont {Engels},
  \citenamefont {Atherton},\ and\ \citenamefont {Hoefer}}]{Engels07}%
  \BibitemOpen
  \bibfield  {author} {\bibinfo {author} {\bibfnamefont {P.}~\bibnamefont
  {Engels}}, \bibinfo {author} {\bibfnamefont {C.}~\bibnamefont {Atherton}},\
  and\ \bibinfo {author} {\bibfnamefont {M.~A.}\ \bibnamefont {Hoefer}},\
  }\bibfield  {title} {\bibinfo {title} {Observation of {Faraday} {Waves} in a
  {Bose}-{Einstein} {Condensate}},\ }\href
  {https://doi.org/10.1103/PhysRevLett.98.095301} {\bibfield  {journal}
  {\bibinfo  {journal} {Physical Review Letters}\ }\textbf {\bibinfo {volume}
  {98}},\ \bibinfo {pages} {095301} (\bibinfo {year} {2007})}\BibitemShut
  {NoStop}%
\bibitem [{\citenamefont {Cominotti}\ \emph {et~al.}(2022)\citenamefont
  {Cominotti}, \citenamefont {Berti}, \citenamefont {Farolfi}, \citenamefont
  {Zenesini}, \citenamefont {Lamporesi}, \citenamefont {Carusotto},
  \citenamefont {Recati},\ and\ \citenamefont {Ferrari}}]{Cominotti22}%
  \BibitemOpen
  \bibfield  {author} {\bibinfo {author} {\bibfnamefont {R.}~\bibnamefont
  {Cominotti}}, \bibinfo {author} {\bibfnamefont {A.}~\bibnamefont {Berti}},
  \bibinfo {author} {\bibfnamefont {A.}~\bibnamefont {Farolfi}}, \bibinfo
  {author} {\bibfnamefont {A.}~\bibnamefont {Zenesini}}, \bibinfo {author}
  {\bibfnamefont {G.}~\bibnamefont {Lamporesi}}, \bibinfo {author}
  {\bibfnamefont {I.}~\bibnamefont {Carusotto}}, \bibinfo {author}
  {\bibfnamefont {A.}~\bibnamefont {Recati}},\ and\ \bibinfo {author}
  {\bibfnamefont {G.}~\bibnamefont {Ferrari}},\ }\bibfield  {title} {\bibinfo
  {title} {Observation of {Massless} and {Massive} {Collective} {Excitations}
  with {Faraday} {Patterns} in a {Two}-{Component} {Superfluid}},\ }\href
  {https://doi.org/10.1103/PhysRevLett.128.210401} {\bibfield  {journal}
  {\bibinfo  {journal} {Physical Review Letters}\ }\textbf {\bibinfo {volume}
  {128}},\ \bibinfo {pages} {210401} (\bibinfo {year} {2022})}\BibitemShut
  {NoStop}%
\bibitem [{\citenamefont {Madison}\ \emph {et~al.}(2001)\citenamefont
  {Madison}, \citenamefont {Chevy}, \citenamefont {Bretin},\ and\ \citenamefont
  {Dalibard}}]{Madison01}%
  \BibitemOpen
  \bibfield  {author} {\bibinfo {author} {\bibfnamefont {K.~W.}\ \bibnamefont
  {Madison}}, \bibinfo {author} {\bibfnamefont {F.}~\bibnamefont {Chevy}},
  \bibinfo {author} {\bibfnamefont {V.}~\bibnamefont {Bretin}},\ and\ \bibinfo
  {author} {\bibfnamefont {J.}~\bibnamefont {Dalibard}},\ }\bibfield  {title}
  {\bibinfo {title} {Stationary {States} of a {Rotating} {Bose}-{Einstein}
  {Condensate}: {Routes} to {Vortex} {Nucleation}},\ }\href
  {https://doi.org/10.1103/PhysRevLett.86.4443} {\bibfield  {journal} {\bibinfo
   {journal} {Physical Review Letters}\ }\textbf {\bibinfo {volume} {86}},\
  \bibinfo {pages} {4443} (\bibinfo {year} {2001})}\BibitemShut {NoStop}%
\bibitem [{\citenamefont {Sinha}\ and\ \citenamefont {Castin}(2001)}]{Sinha01}%
  \BibitemOpen
  \bibfield  {author} {\bibinfo {author} {\bibfnamefont {S.}~\bibnamefont
  {Sinha}}\ and\ \bibinfo {author} {\bibfnamefont {Y.}~\bibnamefont {Castin}},\
  }\bibfield  {title} {\bibinfo {title} {Dynamic {Instability} of a {Rotating}
  {Bose}-{Einstein} {Condensate}},\ }\href
  {https://doi.org/10.1103/PhysRevLett.87.190402} {\bibfield  {journal}
  {\bibinfo  {journal} {Physical Review Letters}\ }\textbf {\bibinfo {volume}
  {87}},\ \bibinfo {pages} {190402} (\bibinfo {year} {2001})}\BibitemShut
  {NoStop}%
\bibitem [{\citenamefont {Nguyen}\ \emph {et~al.}(2017)\citenamefont {Nguyen},
  \citenamefont {Luo},\ and\ \citenamefont {Hulet}}]{Nguyen17}%
  \BibitemOpen
  \bibfield  {author} {\bibinfo {author} {\bibfnamefont {J.~H.~V.}\
  \bibnamefont {Nguyen}}, \bibinfo {author} {\bibfnamefont {D.}~\bibnamefont
  {Luo}},\ and\ \bibinfo {author} {\bibfnamefont {R.~G.}\ \bibnamefont
  {Hulet}},\ }\bibfield  {title} {\bibinfo {title} {Formation of matter-wave
  soliton trains by modulational instability},\ }\href
  {https://doi.org/10.1126/science.aal3220} {\bibfield  {journal} {\bibinfo
  {journal} {Science}\ }\textbf {\bibinfo {volume} {356}},\ \bibinfo {pages}
  {422} (\bibinfo {year} {2017})}\BibitemShut {NoStop}%
\bibitem [{\citenamefont {Nguyen}\ \emph {et~al.}(2019)\citenamefont {Nguyen},
  \citenamefont {Tsatsos}, \citenamefont {Luo}, \citenamefont {Lode},
  \citenamefont {Telles}, \citenamefont {Bagnato},\ and\ \citenamefont
  {Hulet}}]{Nguyen19}%
  \BibitemOpen
  \bibfield  {author} {\bibinfo {author} {\bibfnamefont {J.}~\bibnamefont
  {Nguyen}}, \bibinfo {author} {\bibfnamefont {M.}~\bibnamefont {Tsatsos}},
  \bibinfo {author} {\bibfnamefont {D.}~\bibnamefont {Luo}}, \bibinfo {author}
  {\bibfnamefont {A.}~\bibnamefont {Lode}}, \bibinfo {author} {\bibfnamefont
  {G.}~\bibnamefont {Telles}}, \bibinfo {author} {\bibfnamefont
  {V.}~\bibnamefont {Bagnato}},\ and\ \bibinfo {author} {\bibfnamefont
  {R.}~\bibnamefont {Hulet}},\ }\bibfield  {title} {\bibinfo {title}
  {Parametric {Excitation} of a {Bose}-{Einstein} {Condensate}: {From}
  {Faraday} {Waves} to {Granulation}},\ }\href
  {https://doi.org/10.1103/PhysRevX.9.011052} {\bibfield  {journal} {\bibinfo
  {journal} {Physical Review X}\ }\textbf {\bibinfo {volume} {9}},\ \bibinfo
  {pages} {011052} (\bibinfo {year} {2019})}\BibitemShut {NoStop}%
\bibitem [{\citenamefont {Zhang}\ \emph {et~al.}(2020)\citenamefont {Zhang},
  \citenamefont {Yao}, \citenamefont {Feng}, \citenamefont {Hu},\ and\
  \citenamefont {Chin}}]{Zhang20}%
  \BibitemOpen
  \bibfield  {author} {\bibinfo {author} {\bibfnamefont {Z.}~\bibnamefont
  {Zhang}}, \bibinfo {author} {\bibfnamefont {K.-X.}\ \bibnamefont {Yao}},
  \bibinfo {author} {\bibfnamefont {L.}~\bibnamefont {Feng}}, \bibinfo {author}
  {\bibfnamefont {J.}~\bibnamefont {Hu}},\ and\ \bibinfo {author}
  {\bibfnamefont {C.}~\bibnamefont {Chin}},\ }\bibfield  {title} {\bibinfo
  {title} {Pattern formation in a driven {Bose}–{Einstein} condensate},\
  }\href {https://doi.org/10.1038/s41567-020-0839-3} {\bibfield  {journal}
  {\bibinfo  {journal} {Nature Physics}\ }\textbf {\bibinfo {volume} {16}},\
  \bibinfo {pages} {652} (\bibinfo {year} {2020})}\BibitemShut {NoStop}%
\bibitem [{\citenamefont {Goldman}\ and\ \citenamefont
  {Dalibard}(2014)}]{Goldman14}%
  \BibitemOpen
  \bibfield  {author} {\bibinfo {author} {\bibfnamefont {N.}~\bibnamefont
  {Goldman}}\ and\ \bibinfo {author} {\bibfnamefont {J.}~\bibnamefont
  {Dalibard}},\ }\bibfield  {title} {\bibinfo {title} {{Periodically Driven
  Quantum Systems: Effective Hamiltonians and Engineered Gauge Fields}},\
  }\href {https://doi.org/10.1103/PhysRevX.4.031027} {\bibfield  {journal}
  {\bibinfo  {journal} {Phys. Rev. X}\ }\textbf {\bibinfo {volume} {4}},\
  \bibinfo {pages} {031027} (\bibinfo {year} {2014})}\BibitemShut {NoStop}%
\bibitem [{\citenamefont {Eckardt}(2017)}]{Eckardt17}%
  \BibitemOpen
  \bibfield  {author} {\bibinfo {author} {\bibfnamefont {A.}~\bibnamefont
  {Eckardt}},\ }\bibfield  {title} {\bibinfo {title} {{Colloquium: Atomic
  quantum gases in periodically driven optical lattices}},\ }\href
  {https://doi.org/10.1103/RevModPhys.89.011004} {\bibfield  {journal}
  {\bibinfo  {journal} {Rev. Mod. Phys.}\ }\textbf {\bibinfo {volume} {89}},\
  \bibinfo {pages} {011004} (\bibinfo {year} {2017})}\BibitemShut {NoStop}%
\bibitem [{\citenamefont {Weitenberg}\ and\ \citenamefont
  {Simonet}(2021)}]{Weitenberg21}%
  \BibitemOpen
  \bibfield  {author} {\bibinfo {author} {\bibfnamefont {C.}~\bibnamefont
  {Weitenberg}}\ and\ \bibinfo {author} {\bibfnamefont {J.}~\bibnamefont
  {Simonet}},\ }\bibfield  {title} {\bibinfo {title} {{Tailoring quantum gases
  by Floquet engineering}},\ }\href
  {https://doi.org/10.1038/s41567-021-01316-x} {\bibfield  {journal} {\bibinfo
  {journal} {Nature Physics}\ }\textbf {\bibinfo {volume} {17}},\ \bibinfo
  {pages} {1342} (\bibinfo {year} {2021})}\BibitemShut {NoStop}%
\bibitem [{\citenamefont {Lignier}\ \emph {et~al.}(2007)\citenamefont
  {Lignier}, \citenamefont {Sias}, \citenamefont {Ciampini}, \citenamefont
  {Singh}, \citenamefont {Zenesini}, \citenamefont {Morsch},\ and\
  \citenamefont {Arimondo}}]{Lignier07}%
  \BibitemOpen
  \bibfield  {author} {\bibinfo {author} {\bibfnamefont {H.}~\bibnamefont
  {Lignier}}, \bibinfo {author} {\bibfnamefont {C.}~\bibnamefont {Sias}},
  \bibinfo {author} {\bibfnamefont {D.}~\bibnamefont {Ciampini}}, \bibinfo
  {author} {\bibfnamefont {Y.}~\bibnamefont {Singh}}, \bibinfo {author}
  {\bibfnamefont {A.}~\bibnamefont {Zenesini}}, \bibinfo {author}
  {\bibfnamefont {O.}~\bibnamefont {Morsch}},\ and\ \bibinfo {author}
  {\bibfnamefont {E.}~\bibnamefont {Arimondo}},\ }\bibfield  {title} {\bibinfo
  {title} {{Dynamical Control of Matter-Wave Tunneling in Periodic
  Potentials}},\ }\href {https://doi.org/10.1103/PhysRevLett.99.220403}
  {\bibfield  {journal} {\bibinfo  {journal} {Phys. Rev. Lett.}\ }\textbf
  {\bibinfo {volume} {99}},\ \bibinfo {pages} {220403} (\bibinfo {year}
  {2007})}\BibitemShut {NoStop}%
\bibitem [{\citenamefont {Kierig}\ \emph {et~al.}(2008)\citenamefont {Kierig},
  \citenamefont {Schnorrberger}, \citenamefont {Schietinger}, \citenamefont
  {Tomkovic},\ and\ \citenamefont {Oberthaler}}]{Kierig08}%
  \BibitemOpen
  \bibfield  {author} {\bibinfo {author} {\bibfnamefont {E.}~\bibnamefont
  {Kierig}}, \bibinfo {author} {\bibfnamefont {U.}~\bibnamefont
  {Schnorrberger}}, \bibinfo {author} {\bibfnamefont {A.}~\bibnamefont
  {Schietinger}}, \bibinfo {author} {\bibfnamefont {J.}~\bibnamefont
  {Tomkovic}},\ and\ \bibinfo {author} {\bibfnamefont {M.~K.}\ \bibnamefont
  {Oberthaler}},\ }\bibfield  {title} {\bibinfo {title} {{Single-Particle
  Tunneling in Strongly Driven Double-Well Potentials}},\ }\href
  {https://doi.org/10.1103/PhysRevLett.100.190405} {\bibfield  {journal}
  {\bibinfo  {journal} {Phys. Rev. Lett.}\ }\textbf {\bibinfo {volume} {100}},\
  \bibinfo {pages} {190405} (\bibinfo {year} {2008})}\BibitemShut {NoStop}%
\bibitem [{\citenamefont {Ha}\ \emph {et~al.}(2015)\citenamefont {Ha},
  \citenamefont {Clark}, \citenamefont {Parker}, \citenamefont {Anderson},\
  and\ \citenamefont {Chin}}]{Ha15}%
  \BibitemOpen
  \bibfield  {author} {\bibinfo {author} {\bibfnamefont {L.-C.}\ \bibnamefont
  {Ha}}, \bibinfo {author} {\bibfnamefont {L.~W.}\ \bibnamefont {Clark}},
  \bibinfo {author} {\bibfnamefont {C.~V.}\ \bibnamefont {Parker}}, \bibinfo
  {author} {\bibfnamefont {B.~M.}\ \bibnamefont {Anderson}},\ and\ \bibinfo
  {author} {\bibfnamefont {C.}~\bibnamefont {Chin}},\ }\bibfield  {title}
  {\bibinfo {title} {Roton-{Maxon} {Excitation} {Spectrum} of {Bose}
  {Condensates} in a {Shaken} {Optical} {Lattice}},\ }\href
  {https://doi.org/10.1103/PhysRevLett.114.055301} {\bibfield  {journal}
  {\bibinfo  {journal} {Physical Review Letters}\ }\textbf {\bibinfo {volume}
  {114}},\ \bibinfo {pages} {055301} (\bibinfo {year} {2015})}\BibitemShut
  {NoStop}%
\bibitem [{\citenamefont {Parker}\ \emph {et~al.}(2013)\citenamefont {Parker},
  \citenamefont {Ha},\ and\ \citenamefont {Chin}}]{Parker13}%
  \BibitemOpen
  \bibfield  {author} {\bibinfo {author} {\bibfnamefont {C.~V.}\ \bibnamefont
  {Parker}}, \bibinfo {author} {\bibfnamefont {L.-C.}\ \bibnamefont {Ha}},\
  and\ \bibinfo {author} {\bibfnamefont {C.}~\bibnamefont {Chin}},\ }\bibfield
  {title} {\bibinfo {title} {Direct observation of effective ferromagnetic
  domains of cold atoms in a shaken optical lattice},\ }\href
  {https://doi.org/10.1038/nphys2789} {\bibfield  {journal} {\bibinfo
  {journal} {Nature Physics}\ }\textbf {\bibinfo {volume} {9}},\ \bibinfo
  {pages} {769} (\bibinfo {year} {2013})}\BibitemShut {NoStop}%
\bibitem [{\citenamefont {Feng}\ \emph {et~al.}(2018)\citenamefont {Feng},
  \citenamefont {Clark}, \citenamefont {Gaj},\ and\ \citenamefont
  {Chin}}]{Feng18}%
  \BibitemOpen
  \bibfield  {author} {\bibinfo {author} {\bibfnamefont {L.}~\bibnamefont
  {Feng}}, \bibinfo {author} {\bibfnamefont {L.~W.}\ \bibnamefont {Clark}},
  \bibinfo {author} {\bibfnamefont {A.}~\bibnamefont {Gaj}},\ and\ \bibinfo
  {author} {\bibfnamefont {C.}~\bibnamefont {Chin}},\ }\bibfield  {title}
  {\bibinfo {title} {Coherent inflationary dynamics for {Bose}–{Einstein}
  condensates crossing a quantum critical point},\ }\href
  {https://doi.org/10.1038/s41567-017-0011-x} {\bibfield  {journal} {\bibinfo
  {journal} {Nature Physics}\ }\textbf {\bibinfo {volume} {14}},\ \bibinfo
  {pages} {269} (\bibinfo {year} {2018})}\BibitemShut {NoStop}%
\bibitem [{\citenamefont {Song}\ \emph {et~al.}(2022)\citenamefont {Song},
  \citenamefont {Dutta}, \citenamefont {Bhave}, \citenamefont {Yu},
  \citenamefont {Carter}, \citenamefont {Cooper},\ and\ \citenamefont
  {Schneider}}]{Song22}%
  \BibitemOpen
  \bibfield  {author} {\bibinfo {author} {\bibfnamefont {B.}~\bibnamefont
  {Song}}, \bibinfo {author} {\bibfnamefont {S.}~\bibnamefont {Dutta}},
  \bibinfo {author} {\bibfnamefont {S.}~\bibnamefont {Bhave}}, \bibinfo
  {author} {\bibfnamefont {J.-C.}\ \bibnamefont {Yu}}, \bibinfo {author}
  {\bibfnamefont {E.}~\bibnamefont {Carter}}, \bibinfo {author} {\bibfnamefont
  {N.}~\bibnamefont {Cooper}},\ and\ \bibinfo {author} {\bibfnamefont
  {U.}~\bibnamefont {Schneider}},\ }\bibfield  {title} {\bibinfo {title}
  {Realizing discontinuous quantum phase transitions in a strongly correlated
  driven optical lattice},\ }\href
  {https://doi.org/https://doi.org/10.1038/s41567-021-01476-w} {\bibfield
  {journal} {\bibinfo  {journal} {Nature Physics}\ }\textbf {\bibinfo {volume}
  {18}},\ \bibinfo {pages} {259} (\bibinfo {year} {2022})}\BibitemShut
  {NoStop}%
\bibitem [{\citenamefont {Struck}\ \emph {et~al.}(2011)\citenamefont {Struck},
  \citenamefont {Ölschläger}, \citenamefont {Le~Targat}, \citenamefont
  {Soltan-Panahi}, \citenamefont {Eckardt}, \citenamefont {Lewenstein},
  \citenamefont {Windpassinger},\ and\ \citenamefont {Sengstock}}]{Struck11}%
  \BibitemOpen
  \bibfield  {author} {\bibinfo {author} {\bibfnamefont {J.}~\bibnamefont
  {Struck}}, \bibinfo {author} {\bibfnamefont {C.}~\bibnamefont
  {Ölschläger}}, \bibinfo {author} {\bibfnamefont {R.}~\bibnamefont
  {Le~Targat}}, \bibinfo {author} {\bibfnamefont {P.}~\bibnamefont
  {Soltan-Panahi}}, \bibinfo {author} {\bibfnamefont {A.}~\bibnamefont
  {Eckardt}}, \bibinfo {author} {\bibfnamefont {M.}~\bibnamefont {Lewenstein}},
  \bibinfo {author} {\bibfnamefont {P.}~\bibnamefont {Windpassinger}},\ and\
  \bibinfo {author} {\bibfnamefont {K.}~\bibnamefont {Sengstock}},\ }\bibfield
  {title} {\bibinfo {title} {Quantum {Simulation} of {Frustrated} {Classical}
  {Magnetism} in {Triangular} {Optical} {Lattices}},\ }\href
  {https://doi.org/10.1126/science.1207239} {\bibfield  {journal} {\bibinfo
  {journal} {Science}\ }\textbf {\bibinfo {volume} {333}},\ \bibinfo {pages}
  {996} (\bibinfo {year} {2011})}\BibitemShut {NoStop}%
\bibitem [{\citenamefont {Aidelsburger}\ \emph {et~al.}(2011)\citenamefont
  {Aidelsburger}, \citenamefont {Atala}, \citenamefont {Nascimbène},
  \citenamefont {Trotzky}, \citenamefont {Chen},\ and\ \citenamefont
  {Bloch}}]{Aidelsburger11}%
  \BibitemOpen
  \bibfield  {author} {\bibinfo {author} {\bibfnamefont {M.}~\bibnamefont
  {Aidelsburger}}, \bibinfo {author} {\bibfnamefont {M.}~\bibnamefont {Atala}},
  \bibinfo {author} {\bibfnamefont {S.}~\bibnamefont {Nascimbène}}, \bibinfo
  {author} {\bibfnamefont {S.}~\bibnamefont {Trotzky}}, \bibinfo {author}
  {\bibfnamefont {Y.-A.}\ \bibnamefont {Chen}},\ and\ \bibinfo {author}
  {\bibfnamefont {I.}~\bibnamefont {Bloch}},\ }\bibfield  {title} {\bibinfo
  {title} {Experimental {Realization} of {Strong} {Effective} {Magnetic}
  {Fields} in an {Optical} {Lattice}},\ }\href
  {https://doi.org/10.1103/PhysRevLett.107.255301} {\bibfield  {journal}
  {\bibinfo  {journal} {Physical Review Letters}\ }\textbf {\bibinfo {volume}
  {107}},\ \bibinfo {pages} {255301} (\bibinfo {year} {2011})}\BibitemShut
  {NoStop}%
\bibitem [{\citenamefont {Cooper}\ \emph {et~al.}(2019)\citenamefont {Cooper},
  \citenamefont {Dalibard},\ and\ \citenamefont {Spielman}}]{Cooper19}%
  \BibitemOpen
  \bibfield  {author} {\bibinfo {author} {\bibfnamefont {N.}~\bibnamefont
  {Cooper}}, \bibinfo {author} {\bibfnamefont {J.}~\bibnamefont {Dalibard}},\
  and\ \bibinfo {author} {\bibfnamefont {I.}~\bibnamefont {Spielman}},\
  }\bibfield  {title} {\bibinfo {title} {Topological bands for ultracold
  atoms},\ }\href {https://doi.org/10.1103/RevModPhys.91.015005} {\bibfield
  {journal} {\bibinfo  {journal} {Reviews of Modern Physics}\ }\textbf
  {\bibinfo {volume} {91}},\ \bibinfo {pages} {015005} (\bibinfo {year}
  {2019})}\BibitemShut {NoStop}%
\bibitem [{\citenamefont {Lellouch}\ \emph {et~al.}(2017)\citenamefont
  {Lellouch}, \citenamefont {Bukov}, \citenamefont {Demler},\ and\
  \citenamefont {Goldman}}]{Lellouch17}%
  \BibitemOpen
  \bibfield  {author} {\bibinfo {author} {\bibfnamefont {S.}~\bibnamefont
  {Lellouch}}, \bibinfo {author} {\bibfnamefont {M.}~\bibnamefont {Bukov}},
  \bibinfo {author} {\bibfnamefont {E.}~\bibnamefont {Demler}},\ and\ \bibinfo
  {author} {\bibfnamefont {N.}~\bibnamefont {Goldman}},\ }\bibfield  {title}
  {\bibinfo {title} {{Parametric Instability Rates in Periodically Driven Band
  Systems}},\ }\href {https://doi.org/10.1103/PhysRevX.7.021015} {\bibfield
  {journal} {\bibinfo  {journal} {Phys. Rev. X}\ }\textbf {\bibinfo {volume}
  {7}},\ \bibinfo {pages} {021015} (\bibinfo {year} {2017})}\BibitemShut
  {NoStop}%
\bibitem [{\citenamefont {Boulier}\ \emph {et~al.}(2019)\citenamefont
  {Boulier}, \citenamefont {Maslek}, \citenamefont {Bukov}, \citenamefont
  {Bracamontes}, \citenamefont {Magnan}, \citenamefont {Lellouch},
  \citenamefont {Demler}, \citenamefont {Goldman},\ and\ \citenamefont
  {Porto}}]{Boulier19}%
  \BibitemOpen
  \bibfield  {author} {\bibinfo {author} {\bibfnamefont {T.}~\bibnamefont
  {Boulier}}, \bibinfo {author} {\bibfnamefont {J.}~\bibnamefont {Maslek}},
  \bibinfo {author} {\bibfnamefont {M.}~\bibnamefont {Bukov}}, \bibinfo
  {author} {\bibfnamefont {C.}~\bibnamefont {Bracamontes}}, \bibinfo {author}
  {\bibfnamefont {E.}~\bibnamefont {Magnan}}, \bibinfo {author} {\bibfnamefont
  {S.}~\bibnamefont {Lellouch}}, \bibinfo {author} {\bibfnamefont
  {E.}~\bibnamefont {Demler}}, \bibinfo {author} {\bibfnamefont
  {N.}~\bibnamefont {Goldman}},\ and\ \bibinfo {author} {\bibfnamefont
  {J.}~\bibnamefont {Porto}},\ }\bibfield  {title} {\bibinfo {title}
  {{Parametric Heating in a 2D Periodically Driven Bosonic System: Beyond the
  Weakly Interacting Regime}},\ }\href
  {https://doi.org/https://doi.org/10.1103/PhysRevX.9.011047} {\bibfield
  {journal} {\bibinfo  {journal} {Phys. Rev. X}\ }\textbf {\bibinfo {volume}
  {9}},\ \bibinfo {pages} {011047} (\bibinfo {year} {2019})}\BibitemShut
  {NoStop}%
\bibitem [{\citenamefont {Wintersperger}\ \emph {et~al.}(2020)\citenamefont
  {Wintersperger}, \citenamefont {Bukov}, \citenamefont {N\"ager},
  \citenamefont {Lellouch}, \citenamefont {Demler}, \citenamefont {Schneider},
  \citenamefont {Bloch}, \citenamefont {Goldman},\ and\ \citenamefont
  {Aidelsburger}}]{Wintersperger20}%
  \BibitemOpen
  \bibfield  {author} {\bibinfo {author} {\bibfnamefont {K.}~\bibnamefont
  {Wintersperger}}, \bibinfo {author} {\bibfnamefont {M.}~\bibnamefont
  {Bukov}}, \bibinfo {author} {\bibfnamefont {J.}~\bibnamefont {N\"ager}},
  \bibinfo {author} {\bibfnamefont {S.}~\bibnamefont {Lellouch}}, \bibinfo
  {author} {\bibfnamefont {E.}~\bibnamefont {Demler}}, \bibinfo {author}
  {\bibfnamefont {U.}~\bibnamefont {Schneider}}, \bibinfo {author}
  {\bibfnamefont {I.}~\bibnamefont {Bloch}}, \bibinfo {author} {\bibfnamefont
  {N.}~\bibnamefont {Goldman}},\ and\ \bibinfo {author} {\bibfnamefont
  {M.}~\bibnamefont {Aidelsburger}},\ }\bibfield  {title} {\bibinfo {title}
  {{Parametric Instabilities of Interacting Bosons in Periodically Driven 1D
  Optical Lattices}},\ }\href {https://doi.org/10.1103/PhysRevX.10.011030}
  {\bibfield  {journal} {\bibinfo  {journal} {Phys. Rev. X}\ }\textbf {\bibinfo
  {volume} {10}},\ \bibinfo {pages} {011030} (\bibinfo {year}
  {2020})}\BibitemShut {NoStop}%
\bibitem [{\citenamefont {Rubio-Abadal}\ \emph {et~al.}(2020)\citenamefont
  {Rubio-Abadal}, \citenamefont {Ippoliti}, \citenamefont {Hollerith},
  \citenamefont {Wei}, \citenamefont {Rui}, \citenamefont {Sondhi},
  \citenamefont {Khemani}, \citenamefont {Gross},\ and\ \citenamefont
  {Bloch}}]{Rubio20}%
  \BibitemOpen
  \bibfield  {author} {\bibinfo {author} {\bibfnamefont {A.}~\bibnamefont
  {Rubio-Abadal}}, \bibinfo {author} {\bibfnamefont {M.}~\bibnamefont
  {Ippoliti}}, \bibinfo {author} {\bibfnamefont {S.}~\bibnamefont {Hollerith}},
  \bibinfo {author} {\bibfnamefont {D.}~\bibnamefont {Wei}}, \bibinfo {author}
  {\bibfnamefont {J.}~\bibnamefont {Rui}}, \bibinfo {author} {\bibfnamefont
  {S.~L.}\ \bibnamefont {Sondhi}}, \bibinfo {author} {\bibfnamefont
  {V.}~\bibnamefont {Khemani}}, \bibinfo {author} {\bibfnamefont
  {C.}~\bibnamefont {Gross}},\ and\ \bibinfo {author} {\bibfnamefont
  {I.}~\bibnamefont {Bloch}},\ }\bibfield  {title} {\bibinfo {title} {{Floquet
  Prethermalization in a Bose-Hubbard System}},\ }\href
  {https://doi.org/10.1103/PhysRevX.10.021044} {\bibfield  {journal} {\bibinfo
  {journal} {Phys. Rev. X}\ }\textbf {\bibinfo {volume} {10}},\ \bibinfo
  {pages} {021044} (\bibinfo {year} {2020})}\BibitemShut {NoStop}%
\bibitem [{\citenamefont {Campbell}\ \emph {et~al.}(2006)\citenamefont
  {Campbell}, \citenamefont {Mun}, \citenamefont {Boyd}, \citenamefont
  {Streed}, \citenamefont {Ketterle},\ and\ \citenamefont
  {Pritchard}}]{Campbell06}%
  \BibitemOpen
  \bibfield  {author} {\bibinfo {author} {\bibfnamefont {G.~K.}\ \bibnamefont
  {Campbell}}, \bibinfo {author} {\bibfnamefont {J.}~\bibnamefont {Mun}},
  \bibinfo {author} {\bibfnamefont {M.}~\bibnamefont {Boyd}}, \bibinfo {author}
  {\bibfnamefont {E.~W.}\ \bibnamefont {Streed}}, \bibinfo {author}
  {\bibfnamefont {W.}~\bibnamefont {Ketterle}},\ and\ \bibinfo {author}
  {\bibfnamefont {D.~E.}\ \bibnamefont {Pritchard}},\ }\bibfield  {title}
  {\bibinfo {title} {{Parametric Amplification of Scattered Atom Pairs}},\
  }\href {https://doi.org/10.1103/PhysRevLett.96.020406} {\bibfield  {journal}
  {\bibinfo  {journal} {Phys. Rev. Lett.}\ }\textbf {\bibinfo {volume} {96}},\
  \bibinfo {pages} {020406} (\bibinfo {year} {2006})}\BibitemShut {NoStop}%
\bibitem [{\citenamefont {Galilo}\ \emph {et~al.}(2015)\citenamefont {Galilo},
  \citenamefont {Lee},\ and\ \citenamefont {Barnett}}]{Galilo15}%
  \BibitemOpen
  \bibfield  {author} {\bibinfo {author} {\bibfnamefont {B.}~\bibnamefont
  {Galilo}}, \bibinfo {author} {\bibfnamefont {D.~K.~K.}\ \bibnamefont {Lee}},\
  and\ \bibinfo {author} {\bibfnamefont {R.}~\bibnamefont {Barnett}},\
  }\bibfield  {title} {\bibinfo {title} {{Selective Population of Edge States
  in a 2D Topological Band System}},\ }\href
  {https://doi.org/10.1103/PhysRevLett.115.245302} {\bibfield  {journal}
  {\bibinfo  {journal} {Phys. Rev. Lett.}\ }\textbf {\bibinfo {volume} {115}},\
  \bibinfo {pages} {245302} (\bibinfo {year} {2015})}\BibitemShut {NoStop}%
\bibitem [{\citenamefont {Engelhardt}\ \emph {et~al.}(2016)\citenamefont
  {Engelhardt}, \citenamefont {Benito}, \citenamefont {Platero},\ and\
  \citenamefont {Brandes}}]{Engelhardt16}%
  \BibitemOpen
  \bibfield  {author} {\bibinfo {author} {\bibfnamefont {G.}~\bibnamefont
  {Engelhardt}}, \bibinfo {author} {\bibfnamefont {M.}~\bibnamefont {Benito}},
  \bibinfo {author} {\bibfnamefont {G.}~\bibnamefont {Platero}},\ and\ \bibinfo
  {author} {\bibfnamefont {T.}~\bibnamefont {Brandes}},\ }\bibfield  {title}
  {\bibinfo {title} {{Topological Instabilities in ac-Driven Bosonic
  Systems}},\ }\href {https://doi.org/10.1103/PhysRevLett.117.045302}
  {\bibfield  {journal} {\bibinfo  {journal} {Phys. Rev. Lett.}\ }\textbf
  {\bibinfo {volume} {117}},\ \bibinfo {pages} {045302} (\bibinfo {year}
  {2016})}\BibitemShut {NoStop}%
\bibitem [{\citenamefont {Gemelke}\ \emph {et~al.}(2005)\citenamefont
  {Gemelke}, \citenamefont {Sarajlic}, \citenamefont {Bidel}, \citenamefont
  {Hong},\ and\ \citenamefont {Chu}}]{Gemelke05}%
  \BibitemOpen
  \bibfield  {author} {\bibinfo {author} {\bibfnamefont {N.}~\bibnamefont
  {Gemelke}}, \bibinfo {author} {\bibfnamefont {E.}~\bibnamefont {Sarajlic}},
  \bibinfo {author} {\bibfnamefont {Y.}~\bibnamefont {Bidel}}, \bibinfo
  {author} {\bibfnamefont {S.}~\bibnamefont {Hong}},\ and\ \bibinfo {author}
  {\bibfnamefont {S.}~\bibnamefont {Chu}},\ }\bibfield  {title} {\bibinfo
  {title} {{Parametric Amplification of Matter Waves in Periodically Translated
  Optical Lattices}},\ }\href {https://doi.org/10.1103/PhysRevLett.95.170404}
  {\bibfield  {journal} {\bibinfo  {journal} {Phys. Rev. Lett.}\ }\textbf
  {\bibinfo {volume} {95}},\ \bibinfo {pages} {170404} (\bibinfo {year}
  {2005})}\BibitemShut {NoStop}%
\bibitem [{\citenamefont {Michon}\ \emph {et~al.}(2018)\citenamefont {Michon},
  \citenamefont {Cabrera-Guti{\'{e}}rrez}, \citenamefont {Fortun},
  \citenamefont {Berger}, \citenamefont {Arnal}, \citenamefont {Brunaud},
  \citenamefont {Billy}, \citenamefont {Petitjean}, \citenamefont
  {Schlagheck},\ and\ \citenamefont {Gu{\'{e}}ry-Odelin}}]{Michon18}%
  \BibitemOpen
  \bibfield  {author} {\bibinfo {author} {\bibfnamefont {E.}~\bibnamefont
  {Michon}}, \bibinfo {author} {\bibfnamefont {C.}~\bibnamefont
  {Cabrera-Guti{\'{e}}rrez}}, \bibinfo {author} {\bibfnamefont
  {A.}~\bibnamefont {Fortun}}, \bibinfo {author} {\bibfnamefont
  {M.}~\bibnamefont {Berger}}, \bibinfo {author} {\bibfnamefont
  {M.}~\bibnamefont {Arnal}}, \bibinfo {author} {\bibfnamefont
  {V.}~\bibnamefont {Brunaud}}, \bibinfo {author} {\bibfnamefont
  {J.}~\bibnamefont {Billy}}, \bibinfo {author} {\bibfnamefont
  {C.}~\bibnamefont {Petitjean}}, \bibinfo {author} {\bibfnamefont
  {P.}~\bibnamefont {Schlagheck}},\ and\ \bibinfo {author} {\bibfnamefont
  {D.}~\bibnamefont {Gu{\'{e}}ry-Odelin}},\ }\bibfield  {title} {\bibinfo
  {title} {Phase transition kinetics for a {Bose Einstein} condensate in a
  periodically driven band system},\ }\href
  {https://doi.org/10.1088/1367-2630/aabc3f} {\bibfield  {journal} {\bibinfo
  {journal} {New Journal of Physics}\ }\textbf {\bibinfo {volume} {20}},\
  \bibinfo {pages} {053035} (\bibinfo {year} {2018})}\BibitemShut {NoStop}%
\bibitem [{\citenamefont {Mitchell}\ \emph {et~al.}(2021)\citenamefont
  {Mitchell}, \citenamefont {Di~Carli}, \citenamefont {Sinuco-Le\'on},
  \citenamefont {La~Rooij}, \citenamefont {Kuhr},\ and\ \citenamefont
  {Haller}}]{Mitchell21}%
  \BibitemOpen
  \bibfield  {author} {\bibinfo {author} {\bibfnamefont {M.}~\bibnamefont
  {Mitchell}}, \bibinfo {author} {\bibfnamefont {A.}~\bibnamefont {Di~Carli}},
  \bibinfo {author} {\bibfnamefont {G.}~\bibnamefont {Sinuco-Le\'on}}, \bibinfo
  {author} {\bibfnamefont {A.}~\bibnamefont {La~Rooij}}, \bibinfo {author}
  {\bibfnamefont {S.}~\bibnamefont {Kuhr}},\ and\ \bibinfo {author}
  {\bibfnamefont {E.}~\bibnamefont {Haller}},\ }\bibfield  {title} {\bibinfo
  {title} {{Floquet Solitons and Dynamics of Periodically Driven Matter Waves
  with Negative Effective Mass}},\ }\href
  {https://doi.org/10.1103/PhysRevLett.127.243603} {\bibfield  {journal}
  {\bibinfo  {journal} {Phys. Rev. Lett.}\ }\textbf {\bibinfo {volume} {127}},\
  \bibinfo {pages} {243603} (\bibinfo {year} {2021})}\BibitemShut {NoStop}%
\bibitem [{\citenamefont {Fortun}\ \emph {et~al.}(2016)\citenamefont {Fortun},
  \citenamefont {Cabrera-Guti\'errez}, \citenamefont {Condon}, \citenamefont
  {Michon}, \citenamefont {Billy},\ and\ \citenamefont
  {Gu\'ery-Odelin}}]{Fortun16}%
  \BibitemOpen
  \bibfield  {author} {\bibinfo {author} {\bibfnamefont {A.}~\bibnamefont
  {Fortun}}, \bibinfo {author} {\bibfnamefont {C.}~\bibnamefont
  {Cabrera-Guti\'errez}}, \bibinfo {author} {\bibfnamefont {G.}~\bibnamefont
  {Condon}}, \bibinfo {author} {\bibfnamefont {E.}~\bibnamefont {Michon}},
  \bibinfo {author} {\bibfnamefont {J.}~\bibnamefont {Billy}},\ and\ \bibinfo
  {author} {\bibfnamefont {D.}~\bibnamefont {Gu\'ery-Odelin}},\ }\bibfield
  {title} {\bibinfo {title} {{Direct Tunneling Delay Time Measurement in an
  Optical Lattice}},\ }\href {https://doi.org/10.1103/PhysRevLett.117.010401}
  {\bibfield  {journal} {\bibinfo  {journal} {Phys. Rev. Lett.}\ }\textbf
  {\bibinfo {volume} {117}},\ \bibinfo {pages} {010401} (\bibinfo {year}
  {2016})}\BibitemShut {NoStop}%
\bibitem [{\citenamefont {Greiner}\ \emph
  {et~al.}(2001{\natexlab{a}})\citenamefont {Greiner}, \citenamefont {Bloch},
  \citenamefont {Mandel}, \citenamefont {Hänsch},\ and\ \citenamefont
  {Esslinger}}]{Greiner01}%
  \BibitemOpen
  \bibfield  {author} {\bibinfo {author} {\bibfnamefont {M.}~\bibnamefont
  {Greiner}}, \bibinfo {author} {\bibfnamefont {I.}~\bibnamefont {Bloch}},
  \bibinfo {author} {\bibfnamefont {O.}~\bibnamefont {Mandel}}, \bibinfo
  {author} {\bibfnamefont {T.~W.}\ \bibnamefont {Hänsch}},\ and\ \bibinfo
  {author} {\bibfnamefont {T.}~\bibnamefont {Esslinger}},\ }\bibfield  {title}
  {\bibinfo {title} {Exploring {Phase} {Coherence} in a {2D} {Lattice} of
  {Bose}-{Einstein} {Condensates}},\ }\href
  {https://doi.org/10.1103/PhysRevLett.87.160405} {\bibfield  {journal}
  {\bibinfo  {journal} {Physical Review Letters}\ }\textbf {\bibinfo {volume}
  {87}},\ \bibinfo {pages} {160405} (\bibinfo {year}
  {2001}{\natexlab{a}})}\BibitemShut {NoStop}%
\bibitem [{\citenamefont {Tenart}\ \emph {et~al.}(2020)\citenamefont {Tenart},
  \citenamefont {Carcy}, \citenamefont {Cayla}, \citenamefont {Bourdel},
  \citenamefont {Mancini},\ and\ \citenamefont {Clément}}]{Tenart20}%
  \BibitemOpen
  \bibfield  {author} {\bibinfo {author} {\bibfnamefont {A.}~\bibnamefont
  {Tenart}}, \bibinfo {author} {\bibfnamefont {C.}~\bibnamefont {Carcy}},
  \bibinfo {author} {\bibfnamefont {H.}~\bibnamefont {Cayla}}, \bibinfo
  {author} {\bibfnamefont {T.}~\bibnamefont {Bourdel}}, \bibinfo {author}
  {\bibfnamefont {M.}~\bibnamefont {Mancini}},\ and\ \bibinfo {author}
  {\bibfnamefont {D.}~\bibnamefont {Clément}},\ }\bibfield  {title} {\bibinfo
  {title} {Two-body collisions in the time-of-flight dynamics of lattice {Bose}
  superfluids},\ }\href {https://doi.org/10.1103/PhysRevResearch.2.013017}
  {\bibfield  {journal} {\bibinfo  {journal} {Physical Review Research}\
  }\textbf {\bibinfo {volume} {2}},\ \bibinfo {pages} {013017} (\bibinfo {year}
  {2020})}\BibitemShut {NoStop}%
\bibitem [{\citenamefont {Chatelain}\ \emph {et~al.}(2020)\citenamefont
  {Chatelain}, \citenamefont {Dupont}, \citenamefont {Arnal}, \citenamefont
  {Brunaud}, \citenamefont {Billy}, \citenamefont {Peaudecerf}, \citenamefont
  {Schlagheck},\ and\ \citenamefont {Gu{\'{e}}ry-Odelin}}]{Chatelain_2020}%
  \BibitemOpen
  \bibfield  {author} {\bibinfo {author} {\bibfnamefont {G.}~\bibnamefont
  {Chatelain}}, \bibinfo {author} {\bibfnamefont {N.}~\bibnamefont {Dupont}},
  \bibinfo {author} {\bibfnamefont {M.}~\bibnamefont {Arnal}}, \bibinfo
  {author} {\bibfnamefont {V.}~\bibnamefont {Brunaud}}, \bibinfo {author}
  {\bibfnamefont {J.}~\bibnamefont {Billy}}, \bibinfo {author} {\bibfnamefont
  {B.}~\bibnamefont {Peaudecerf}}, \bibinfo {author} {\bibfnamefont
  {P.}~\bibnamefont {Schlagheck}},\ and\ \bibinfo {author} {\bibfnamefont
  {D.}~\bibnamefont {Gu{\'{e}}ry-Odelin}},\ }\bibfield  {title} {\bibinfo
  {title} {Observation and control of quantized scattering halos},\ }\href
  {https://doi.org/10.1088/1367-2630/abcf6a} {\bibfield  {journal} {\bibinfo
  {journal} {New Journal of Physics}\ }\textbf {\bibinfo {volume} {22}},\
  \bibinfo {pages} {123032} (\bibinfo {year} {2020})}\BibitemShut {NoStop}%
\bibitem [{\citenamefont {Steel}\ \emph {et~al.}(1998)\citenamefont {Steel},
  \citenamefont {Olsen}, \citenamefont {Plimak}, \citenamefont {Drummond},
  \citenamefont {Tan}, \citenamefont {Collett}, \citenamefont {Walls},\ and\
  \citenamefont {Graham}}]{TW1}%
  \BibitemOpen
  \bibfield  {author} {\bibinfo {author} {\bibfnamefont {M.~J.}\ \bibnamefont
  {Steel}}, \bibinfo {author} {\bibfnamefont {M.~K.}\ \bibnamefont {Olsen}},
  \bibinfo {author} {\bibfnamefont {L.~I.}\ \bibnamefont {Plimak}}, \bibinfo
  {author} {\bibfnamefont {P.~D.}\ \bibnamefont {Drummond}}, \bibinfo {author}
  {\bibfnamefont {S.~M.}\ \bibnamefont {Tan}}, \bibinfo {author} {\bibfnamefont
  {M.~J.}\ \bibnamefont {Collett}}, \bibinfo {author} {\bibfnamefont {D.~F.}\
  \bibnamefont {Walls}},\ and\ \bibinfo {author} {\bibfnamefont
  {R.}~\bibnamefont {Graham}},\ }\bibfield  {title} {\bibinfo {title}
  {Dynamical quantum noise in trapped {Bose-Einstein} condensates},\ }\href
  {https://doi.org/10.1103/PhysRevA.58.4824} {\bibfield  {journal} {\bibinfo
  {journal} {Phys. Rev. A}\ }\textbf {\bibinfo {volume} {58}},\ \bibinfo
  {pages} {4824} (\bibinfo {year} {1998})}\BibitemShut {NoStop}%
\bibitem [{\citenamefont {Sinatra}\ \emph {et~al.}(2002)\citenamefont
  {Sinatra}, \citenamefont {Lobo},\ and\ \citenamefont {Castin}}]{TW2}%
  \BibitemOpen
  \bibfield  {author} {\bibinfo {author} {\bibfnamefont {A.}~\bibnamefont
  {Sinatra}}, \bibinfo {author} {\bibfnamefont {C.}~\bibnamefont {Lobo}},\ and\
  \bibinfo {author} {\bibfnamefont {Y.}~\bibnamefont {Castin}},\ }\bibfield
  {title} {\bibinfo {title} {{The truncated Wigner method for Bose-condensed
  gases: limits of validity and applications}},\ }\href
  {https://doi.org/10.1088/0953-4075/35/17/301} {\bibfield  {journal} {\bibinfo
   {journal} {Journal of Physics B: Atomic, Molecular and Optical Physics}\
  }\textbf {\bibinfo {volume} {35}},\ \bibinfo {pages} {3599} (\bibinfo {year}
  {2002})}\BibitemShut {NoStop}%
\bibitem [{\citenamefont {Polkovnikov}(2010)}]{TW3}%
  \BibitemOpen
  \bibfield  {author} {\bibinfo {author} {\bibfnamefont {A.}~\bibnamefont
  {Polkovnikov}},\ }\bibfield  {title} {\bibinfo {title} {Phase space
  representation of quantum dynamics},\ }\href
  {https://doi.org/https://doi.org/10.1016/j.aop.2010.02.006} {\bibfield
  {journal} {\bibinfo  {journal} {Annals of Physics}\ }\textbf {\bibinfo
  {volume} {325}},\ \bibinfo {pages} {1790} (\bibinfo {year}
  {2010})}\BibitemShut {NoStop}%
\bibitem [{\citenamefont {Dujardin}\ \emph {et~al.}(2015)\citenamefont
  {Dujardin}, \citenamefont {Engl}, \citenamefont {Urbina},\ and\ \citenamefont
  {Schlagheck}}]{TW4}%
  \BibitemOpen
  \bibfield  {author} {\bibinfo {author} {\bibfnamefont {J.}~\bibnamefont
  {Dujardin}}, \bibinfo {author} {\bibfnamefont {T.}~\bibnamefont {Engl}},
  \bibinfo {author} {\bibfnamefont {J.~D.}\ \bibnamefont {Urbina}},\ and\
  \bibinfo {author} {\bibfnamefont {P.}~\bibnamefont {Schlagheck}},\ }\bibfield
   {title} {\bibinfo {title} {{Describing many-body bosonic waveguide
  scattering with the truncated Wigner method}},\ }\href
  {https://doi.org/https://doi.org/10.1002/andp.201500113} {\bibfield
  {journal} {\bibinfo  {journal} {Annalen der Physik}\ }\textbf {\bibinfo
  {volume} {527}},\ \bibinfo {pages} {629} (\bibinfo {year}
  {2015})}\BibitemShut {NoStop}%
\bibitem [{\citenamefont {Asteria}\ \emph {et~al.}(2021)\citenamefont
  {Asteria}, \citenamefont {Zahn}, \citenamefont {Kosch}, \citenamefont
  {Sengstock},\ and\ \citenamefont {Weitenberg}}]{Asteria21}%
  \BibitemOpen
  \bibfield  {author} {\bibinfo {author} {\bibfnamefont {L.}~\bibnamefont
  {Asteria}}, \bibinfo {author} {\bibfnamefont {H.~P.}\ \bibnamefont {Zahn}},
  \bibinfo {author} {\bibfnamefont {M.~N.}\ \bibnamefont {Kosch}}, \bibinfo
  {author} {\bibfnamefont {K.}~\bibnamefont {Sengstock}},\ and\ \bibinfo
  {author} {\bibfnamefont {C.}~\bibnamefont {Weitenberg}},\ }\bibfield  {title}
  {\bibinfo {title} {Quantum gas magnifier for sub-lattice-resolved imaging of
  {3D} quantum systems},\ }\href {https://doi.org/10.1038/s41586-021-04011-2}
  {\bibfield  {journal} {\bibinfo  {journal} {Nature}\ }\textbf {\bibinfo
  {volume} {599}},\ \bibinfo {pages} {571} (\bibinfo {year}
  {2021})}\BibitemShut {NoStop}%
\bibitem [{\citenamefont {Zahn}\ \emph {et~al.}(2022)\citenamefont {Zahn},
  \citenamefont {Singh}, \citenamefont {Kosch}, \citenamefont {Asteria},
  \citenamefont {Freystatzky}, \citenamefont {Sengstock}, \citenamefont
  {Mathey},\ and\ \citenamefont {Weitenberg}}]{Zahn22}%
  \BibitemOpen
  \bibfield  {author} {\bibinfo {author} {\bibfnamefont {H.}~\bibnamefont
  {Zahn}}, \bibinfo {author} {\bibfnamefont {V.}~\bibnamefont {Singh}},
  \bibinfo {author} {\bibfnamefont {M.}~\bibnamefont {Kosch}}, \bibinfo
  {author} {\bibfnamefont {L.}~\bibnamefont {Asteria}}, \bibinfo {author}
  {\bibfnamefont {L.}~\bibnamefont {Freystatzky}}, \bibinfo {author}
  {\bibfnamefont {K.}~\bibnamefont {Sengstock}}, \bibinfo {author}
  {\bibfnamefont {L.}~\bibnamefont {Mathey}},\ and\ \bibinfo {author}
  {\bibfnamefont {C.}~\bibnamefont {Weitenberg}},\ }\bibfield  {title}
  {\bibinfo {title} {Formation of {Spontaneous} {Density}-{Wave} {Patterns} in
  dc {Driven} {Lattices}},\ }\href {https://doi.org/10.1103/PhysRevX.12.021014}
  {\bibfield  {journal} {\bibinfo  {journal} {Physical Review X}\ }\textbf
  {\bibinfo {volume} {12}},\ \bibinfo {pages} {021014} (\bibinfo {year}
  {2022})}\BibitemShut {NoStop}%
\bibitem [{\citenamefont {Castin}\ and\ \citenamefont
  {Dum}(1998)}]{castindum98}%
  \BibitemOpen
  \bibfield  {author} {\bibinfo {author} {\bibfnamefont {Y.}~\bibnamefont
  {Castin}}\ and\ \bibinfo {author} {\bibfnamefont {R.}~\bibnamefont {Dum}},\
  }\bibfield  {title} {\bibinfo {title} {Low-temperature {Bose-Einstein}
  condensates in time-dependent traps: Beyond the u(1) symmetry-breaking
  approach},\ }\href {https://doi.org/https://doi.org/10.1103/PhysRevA.57.3008}
  {\bibfield  {journal} {\bibinfo  {journal} {Phys. Rev. A}\ }\textbf {\bibinfo
  {volume} {57}},\ \bibinfo {pages} {3008} (\bibinfo {year}
  {1998})}\BibitemShut {NoStop}%
\bibitem [{\citenamefont {Paul}\ \emph {et~al.}(2007)\citenamefont {Paul},
  \citenamefont {Hartung}, \citenamefont {Richter},\ and\ \citenamefont
  {Schlagheck}}]{Paul07}%
  \BibitemOpen
  \bibfield  {author} {\bibinfo {author} {\bibfnamefont {T.}~\bibnamefont
  {Paul}}, \bibinfo {author} {\bibfnamefont {M.}~\bibnamefont {Hartung}},
  \bibinfo {author} {\bibfnamefont {K.}~\bibnamefont {Richter}},\ and\ \bibinfo
  {author} {\bibfnamefont {P.}~\bibnamefont {Schlagheck}},\ }\bibfield  {title}
  {\bibinfo {title} {Nonlinear transport of {Bose-Einstein} condensates through
  mesoscopic waveguides},\ }\href
  {https://doi.org/https://doi.org/10.1103/PhysRevA.76.063605} {\bibfield
  {journal} {\bibinfo  {journal} {Phys. Rev. A}\ }\textbf {\bibinfo {volume}
  {76}},\ \bibinfo {pages} {063605} (\bibinfo {year} {2007})}\BibitemShut
  {NoStop}%
\bibitem [{\citenamefont {Wanzenb\"ock}\ \emph {et~al.}(2021)\citenamefont
  {Wanzenb\"ock}, \citenamefont {Donsa}, \citenamefont {Hofst\"atter},
  \citenamefont {Koch}, \citenamefont {Schlagheck},\ and\ \citenamefont
  {B\ifmmode~\check{r}\else \v{r}\fi{}ezinov\'a}}]{Wan21PRA}%
  \BibitemOpen
  \bibfield  {author} {\bibinfo {author} {\bibfnamefont {R.}~\bibnamefont
  {Wanzenb\"ock}}, \bibinfo {author} {\bibfnamefont {S.}~\bibnamefont {Donsa}},
  \bibinfo {author} {\bibfnamefont {H.}~\bibnamefont {Hofst\"atter}}, \bibinfo
  {author} {\bibfnamefont {O.}~\bibnamefont {Koch}}, \bibinfo {author}
  {\bibfnamefont {P.}~\bibnamefont {Schlagheck}},\ and\ \bibinfo {author}
  {\bibfnamefont {I.}~\bibnamefont {B\ifmmode~\check{r}\else
  \v{r}\fi{}ezinov\'a}},\ }\bibfield  {title} {\bibinfo {title} {Chaos-induced
  loss of coherence of a {Bose-Einstein} condensate},\ }\href
  {https://doi.org/10.1103/PhysRevA.103.023336} {\bibfield  {journal} {\bibinfo
   {journal} {Phys. Rev. A}\ }\textbf {\bibinfo {volume} {103}},\ \bibinfo
  {pages} {023336} (\bibinfo {year} {2021})}\BibitemShut {NoStop}%
\bibitem [{\citenamefont {Penrose}\ and\ \citenamefont
  {Onsager}(1956)}]{PenOns56PR}%
  \BibitemOpen
  \bibfield  {author} {\bibinfo {author} {\bibfnamefont {O.}~\bibnamefont
  {Penrose}}\ and\ \bibinfo {author} {\bibfnamefont {L.}~\bibnamefont
  {Onsager}},\ }\bibfield  {title} {\bibinfo {title} {{Bose-Einstein
  Condensation and Liquid Helium}},\ }\href
  {https://doi.org/10.1103/PhysRev.104.576} {\bibfield  {journal} {\bibinfo
  {journal} {Phys. Rev.}\ }\textbf {\bibinfo {volume} {104}},\ \bibinfo {pages}
  {576} (\bibinfo {year} {1956})}\BibitemShut {NoStop}%
\bibitem [{\citenamefont {Greiner}\ \emph
  {et~al.}(2001{\natexlab{b}})\citenamefont {Greiner}, \citenamefont {Bloch},
  \citenamefont {Mandel}, \citenamefont {H\"ansch},\ and\ \citenamefont
  {Esslinger}}]{bandmapping}%
  \BibitemOpen
  \bibfield  {author} {\bibinfo {author} {\bibfnamefont {M.}~\bibnamefont
  {Greiner}}, \bibinfo {author} {\bibfnamefont {I.}~\bibnamefont {Bloch}},
  \bibinfo {author} {\bibfnamefont {O.}~\bibnamefont {Mandel}}, \bibinfo
  {author} {\bibfnamefont {T.~W.}\ \bibnamefont {H\"ansch}},\ and\ \bibinfo
  {author} {\bibfnamefont {T.}~\bibnamefont {Esslinger}},\ }\bibfield  {title}
  {\bibinfo {title} {{Exploring Phase Coherence in a 2D Lattice of
  Bose-Einstein Condensates}},\ }\href
  {https://doi.org/10.1103/PhysRevLett.87.160405} {\bibfield  {journal}
  {\bibinfo  {journal} {Phys. Rev. Lett.}\ }\textbf {\bibinfo {volume} {87}},\
  \bibinfo {pages} {160405} (\bibinfo {year} {2001}{\natexlab{b}})}\BibitemShut
  {NoStop}%
\end{thebibliography}%

\begin{appendix}
\section*{Methods}

\subsection{Production of BECs}
\label{appen:setup}

In our experiment, approximately $2\cdot 10^9$ atoms from a 3D magneto-optical trap are initially loaded into a magnetic quadrupole. The quadrupole gradient is then ramped up to $1.8\,\mathrm{T/m}$ to allow for microwave evaporation. After evaporation over $10\,\mathrm{s}$ the temperature of the atom cloud is decreased from $300\,\mathrm{\mu K}$ to $30\,\mathrm{\mu K}$. Subsequently, the atoms are transferred to a crossed dipole trap, made of two $1064\,\mathrm{nm}$ laser beams, with waist $45\,\mathrm{\mu m}$ and maximum power $4\,\mathrm{W}$, crossing in the horizontal plane with a $16^\circ$ angle, with one beam aligned on the lattice axis. After this transfer and throughout the experiment, a magnetic gradient is maintained for gravity compensation. The evaporation in this final dipole trap yields a pure BEC of up to $5\cdot10^5$ atoms in the low-field-seeker state $\ket{F=1,m_F=-1}$.

\subsection{Tight-binding effective model}
\label{appen:model}

\subsubsection{Model and instability}

We model the resonant coupling between bands at $q\neq 0$ in the modulated lattice by an effective tight-binding model with two coupled bands that reproduces the main features of a typical Floquet spectrum, described by the effective Hamiltonian :

\begin{align}
\hat{H}_\mathrm{eff}=&\hat{H}_0+\hat{H}_\mathrm{int} \nonumber \label{eq:hamiltonian}\\ 
\hat{H}_0=&-J_0\sum_\ell \hat{a}_{\ell+1}^\dagger\hat{a}_\ell+\hat{a}_{\ell}^\dagger\hat{a}_{\ell+1}-J_1\sum_\ell \hat{b}_{\ell+1}^\dagger\hat{b}_\ell+\hat{b}_{\ell}^\dagger\hat{b}_{\ell+1}\\
&+E_b\sum_\ell \hat{b}_{\ell}^\dagger\hat{b}_\ell+W\sum_\ell \hat{b}_{\ell}^\dagger\hat{a}_\ell+\hat{a}_{\ell}^\dagger\hat{b}_{\ell} \nonumber\\
\hat{H}_\mathrm{int}=&\frac{U}{2}\sum_\ell \hat{a}_{\ell}^{\dagger 2}\hat{a}_{\ell}^{2},\nonumber
\end{align}
where $\hat{a}_\ell$ (resp. $\hat{b}_\ell$) are the annihilation operators for band $0$ (resp. 1) on site $\ell$ of the one-dimensional lattice, $J_{0,1}$ are the tunneling amplitudes for the two bands ($J_0>0$ and $|J_0|<|J_1|$), $E_b$ is an energy offset for band $1$, $W$ is a coupling amplitude, and $U$ is an effective on-site interaction energy (see Appendix~\ref{appen:interaction}).

The condensate is considered initially in the ground mode of band 0 with $\langle \hat{a}_\ell \rangle=\sqrt{n}$, and associated chemical potential $\mu=-2J_0+nU$. We then study the stability of this initial condensate due to the interaction term, through a perturbative Bogolubov treatment. 
The coupled-band Hamiltonian $\hat{H}_0$ describes two hybridized energy bands $u$ and $v$ with energies $E_{u,v}(q)$, as a function of quasi-momentum $q$:

\begin{align}
E_u(q)&=\left[E_0(q)\cos^2\left(\frac{\theta}{2}\right)+E_1(q)\sin^2\left(\frac{\theta}{2}\right)\right]+W\sin(\theta),\nonumber\\
E_v(q)&=\left[E_0(q)\sin^2\left(\frac{\theta}{2}\right)+E_1(q)\cos^2\left(\frac{\theta}{2}\right)\right]-W\sin(\theta).\nonumber
\end{align}
where $E_{0}(q)=-2J_{0}\cos(q)$ and $E_{1}(q)=-2J_{1}\cos(q)+E_b$ are the energies of the uncoupled bands, and $\theta(q)$ is the mixing angle defined by:

\begin{equation}
\tan(\theta(q))=\frac{2W}{(E_0(q)-E_1(q))},\; 0\leq\theta<\pi.\nonumber
\end{equation}

We write the fields orthogonal to the condensate mode as :

\begin{align}
\delta\hat{\psi}^\perp_{a,\ell}&=\int^\pi_{-\pi}\frac{e^{iq\ell}}{\sqrt{2\pi}}\hat{\Lambda}_{0,q}\,\mathrm{d}q \nonumber\\
\delta\hat{\psi}^\perp_{b,\ell}&=\int^\pi_{-\pi}\frac{e^{iq\ell}}{\sqrt{2\pi}}\hat{\Lambda}_{1,q}\,\mathrm{d}q \nonumber
\end{align}
where we have introduced the number conserving operators $\hat{\Lambda}_{0,q}$ (resp. $\hat{\Lambda}_{1,q}$) which describe the transfer of an atom from the mode at momentum $q$ of band 0 (resp. band 1) to the condensate mode~\cite{castindum98}. 

The quadratic part of the expansion of $\hat{H}_\mathrm{eff}$ in the fields $\hat{\Lambda}$ yields a set of linearized evolution equations, which can be summarized as

\begin{align}
\label{eq:stability}
i\hbar \frac{\mathrm{d}}{\mathrm{d}t}\begin{pmatrix}\hat{\Lambda}_{0,q}\\\hat{\Lambda}_{1,q}\\\hat{\Lambda}^\dagger_{0,-q}\\\hat{\Lambda}^\dagger_{1,-q}\end{pmatrix}&=\mathcal{L}_{q}\begin{pmatrix}\hat{\Lambda}_{0,q}\\\hat{\Lambda}_{1,q}\\\hat{\Lambda}^\dagger_{0,-q}\\\hat{\Lambda}^\dagger_{1,-q}\end{pmatrix}
\end{align}
where $\mathcal{L}_{q}$ is block matrix made of the 2-by-2 matrices $\hat{A}_q$ and $\hat{B}_q$, with definitions:

\begin{align*}
\mathcal{L}_{q}&=\begin{pmatrix}\hat{A}_q & \hat{B}_q\\ -\hat{B}_q & -\hat{A}_q\end{pmatrix}\\
\hat{A}_q&=\begin{pmatrix}E_0(q)-\mu+2nU & W\\ W & E_1(q)-\mu\end{pmatrix}\\
\hat{B}_q&=\begin{pmatrix}nU & 0\\  0& 0\end{pmatrix}\\
\end{align*}

The modes at $q$ are stable if the eigenvalues of the matrix $\mathcal{L}_q$ are all real. In practice we may search for the largest imaginary part among all eigenvalues to characterize instability.

\begin{figure}[ht!]
	\begin{center}
		\includegraphics[width=0.66\columnwidth]{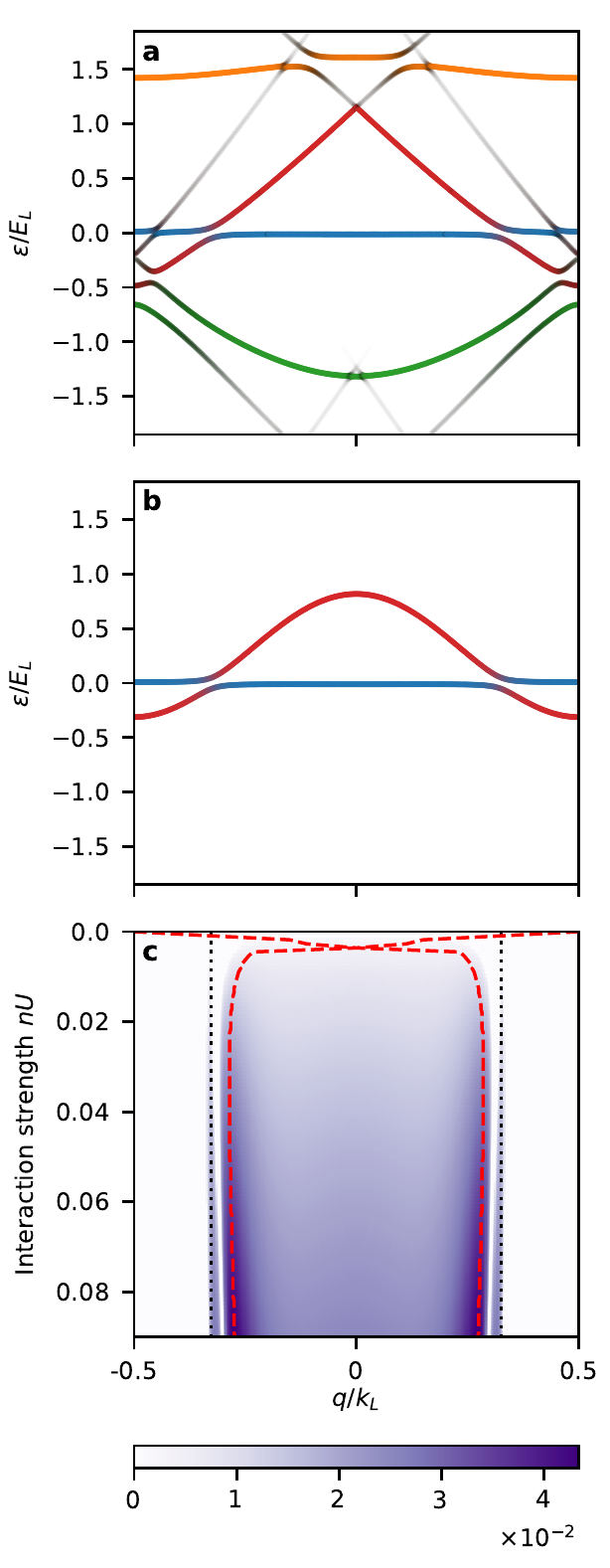}
		\caption{\textbf{Tight-binding model of the instability.} \textbf{(a)} Floquet spectrum of the modulated system for parameters $s_0=3.4$, $\nu=30\,\mathrm{kHz}$ and $\varphi_0=20^\circ$. Overlap with the states of the static lattice bands is color-coded with blue, orange, green and red corresponding respectively to the first 4 bands (\textit{s} to \textit{f}). \textbf{(b)} Spectrum of the model Hamiltonian $H_0$ (see equation \ref{eq:hamiltonian}) with two coupled, tightly-bound bands. Adjusted parameters (see text) are (in units of $E_\mathrm{L}$) $J_0=0.0021$, $J_1=-0.2796$, $E_b=0.2593$ and $W=0.051$ (a global offset is applied to match the Floquet spectrum in \textbf{(a)}). \textbf{(c,d)} Maximum instability exponent of the Bogolubov matrix (\ref{eq:stability}), as a function of quasi-momentum and the interaction parameter $nU$. The position of the band crossing and the maximum exponent over the Brillouin zone are plotted in dotted black and dashed red lines respectively.}
		\label{appen:fig:TBmodel}
	\end{center}
\end{figure}

\paragraph*{Illustration.}
We illustrate the model with conditions similar to those of Fig.\ref{fig:figure1} at a frequency $\nu=30\,\mathrm{kHz}$. We have represented the Floquet spectrum of the modulated lattice (Fig.\ref{appen:fig:TBmodel}{\bf (a)}), and the model spectrum (Fig.\ref{appen:fig:TBmodel}{\bf (b)}) of Hamiltonian $\hat{H}_0$ (see equation \ref{eq:hamiltonian}), with parameters adjusted so that the coupled bands best reproduce the avoided crossings between the $s$ and $d$ bands in the Floquet system (in units of $E_\mathrm{L}$, $J_0=0.0021$, $J_1=-0.2796$, $E_b=0.2593$ and $W=0.051$). In Fig.\ref{appen:fig:TBmodel}{\bf (c)} we plot the instability exponent defined as the absolute value of the largest imaginary part among the four eigenvalues, for a varying value of the interaction parameter $nU$ (see Section \ref{appen:interaction}). Two narrow regions of $q$ in the vicinity of the avoided crossings lead to pure imaginary eigenvalues, with maximally unstable modes.

\subsubsection{Onset of correlations from the instability}

Based on the previous mode decomposition, we can write a general expression for the elements of the reduced one-body density matrix :

\begin{align}
\langle \hat{a}_l^\dagger(t) \hat{a}_{l'}(t) \rangle =n + \int_{-\pi}^{\pi} \frac{ e^{iq (l' - l)}}{2\pi}\langle\hat{\Lambda}^\dagger_{0,q}(t)\hat{\Lambda}_{0,q}(t)\rangle \,\mathrm{d}q
  \label{eq:1bdm}
\end{align}

Due to the symmetries of the Bogolubov matrix $\mathcal{L}_q$, and near the maximum of the instability exponent, its eigenvalues come in pairs of opposite real and imaginary values, which we denote $\{\omega_q,-\omega_q,i\lambda_q, -i\lambda_q\}$ with the convention $\omega_q,\lambda_q>0$. Using the fact that none of the modes with $q \neq 0$ is initially populated, the expression of the average value in Equation~(\ref{eq:1bdm}) can be obtained and is approximately equal to:
\begin{equation}
\langle\hat{\Lambda}^\dagger_{0,q}(t)\hat{\Lambda}_{0,q}(t)\rangle\simeq |u_q|^2(1+|v_q|^2) e^{2\lambda_q t}
\end{equation}
keeping the exponentially diverging terms only, where the coefficients $u_q,v_q$ are the coefficients of the eigenvector of $\mathcal{L}_q$ for the eigenvalue $i\lambda_q$, which is generally of the form $(u_q,v_q,iu_q^*,iv_q^*)^T$.

Let us now consider the vicinity of a maximum of the instability exponent $\lambda_q$, near some $q = q^* > 0$. We have 
\begin{equation}
  \lambda_q \simeq \lambda_*-\frac{\lambda''}{2}(q-q^*)^2+O\left((q-q^*)^3\right)
\end{equation}
with $\lambda_*>0$.
Due to the symmetry of the band structure, the same behavior arises near $q = - q^*$,
\begin{equation}
  \lambda_q \simeq \lambda_*-\frac{\lambda''}{2}(q+q^*)^2+O\left((q+q^*)^3\right)
\end{equation}

We can then evaluate Eq.~\eqref{eq:1bdm} with the saddle-point approximation. This yields the estimate
\begin{equation}
  \langle \hat{a}_l^\dagger(t) \hat{a}_{l'}(t) \rangle \simeq n + 2 n^*(t)
  e^{- (l - l')^2/\Delta^2(t)} \cos[(l - l') q^*]
\end{equation}
with
\begin{equation}
  \Delta(t) = 2\sqrt{\lambda''t} \,, \label{eq:Delta}
\end{equation}
with the time-dependent population
\begin{equation}
  n^*(t) = \frac{|u_{q^*}|^2(1+|v_{q^*}|^2)}{\sqrt{\pi}\Delta(t)} e^{2 \lambda_* t}
\end{equation} 
of excitations near the modes of momentum $q^*$. This implicitly assumes that this population stays much smaller than the
remaining condensate population $n$ at all times.

A very similar result is obtained for the density-density correlation function.
Defining the site population operator $\hat{n}_l = \hat{a}_l^\dagger \hat{a}_l$
and the mean site occupancy
$\bar{n} = \langle \hat{n}_l \rangle = n + 2 n^*$, and using $n^*(t) \ll n$,
we obtain, to first order in $n^*$
\begin{align}
  g^{(2)}(l - l')-1 & = \frac{\langle \hat{n}_l(t) \hat{n}_{l'}(t) \rangle}
  {\langle \hat{n}_l(t) \rangle \langle \hat{n}_{l'}(t) \rangle} - 1 \nonumber \\
 & \simeq  4 \frac{n n^*(t)}{\bar{n}^2} e^{- (l - l')^2/\Delta^2(t)}
  \cos[(l - l') q^*] - \frac{\delta_{l l'}}{\bar{n}} \,.
\end{align}

We therefore have a normalized coherence
$g^{(1)}(l-l') = |\langle \hat{a}_l^\dagger(t) \hat{a}_{l'}(t) \rangle| / \bar{n}$ and normalized correlation $g^{(2)}(l - l')$ that spatially oscillate with the period $d^* = 2\pi/q^*$ (in dimensionless lattice units) and decay to $1$ on the characteristic scale $\Delta(t)$. 

A key parameter for this decay scale is therefore the sharpness of the instability peak (described by the second derivative $\lambda^{''}$). It is the existence of the sharp features in the coupled-band system that allow for an extended order to appear in the system.
While the decay scale $\Delta(t) \propto t^{1/2}$ can, in principle, also be enhanced by increasing
the evolution time $t$, letting the system evolve for too long a time leads
to a breakdown of the Bogolubov approximation, entailing secondary
atom-atom scattering processes and a global loss of coherence.

\subsection{Estimation of the interaction parameter}
\label{appen:interaction}

The interaction parameter $nU$ in the effective 1D model has to take into account the fact that the real system is a 1D lattice of pancakes of atoms. In order to account for weakly populated sites, where the interaction can be described perturbatively, and strongly populated sites, with a transverse Thomas-Fermi profile, we use a heuristic interpolation formula~\cite{Paul07,Michon18} for the interaction energy $U$ on site $\ell$:

\begin{equation}
U_\ell=\frac{2\hbar\omega_\perp a_\mathrm{s}/(\sqrt{2\pi} a_0)}{\sqrt{1+4n_\ell a_\mathrm{s}/(\sqrt{2\pi} a_0)}},
\end{equation}
where $n_\ell$ is the number of atoms on site $\ell$, $\omega_\perp$ is the geometrical average transverse frequency ($\omega_\perp/(2\pi)=67\,\mathrm{Hz}$), $a_\mathrm{s}$ is the scattering length of $^{87}\mathrm{Rb}$ ($a_\mathrm{s}\simeq5.3\,\mathrm{nm}$), and $a_0$ is the characteristic size of the ground state of the lattice potential well in the harmonic approximation: $a_0=\sqrt{\hbar^2/(mE_\mathrm{L}\sqrt{s})}=ds^{-1/4}/(\pi\sqrt{2})\simeq86\,\mathrm{nm}$.

Within this approximation, we can estimate the maximum value of the interaction parameter $nU$. Taking an initial Thomas-Fermi profile for the BEC in the dipole trap with frequencies
$(\omega_{x'}, \omega_{y'}, \omega_z) = 2\pi\times(10.4, 66, 68)$ Hz, and a total number of atoms $N=5\times 10^5$, we estimate that the number of atoms loaded in the central site of the lattice ($\ell=0$) is $n_0\simeq4.4\times10^3$. We can then compute the maximum value $n_0U_0/E_\dL$:
\begin{equation}
\frac{n_0 U_0}{E_\mathrm{L}}\simeq\frac{\hbar\omega_\perp}{E_\mathrm{L}}\sqrt{\frac{a_\mathrm{s}}{\sqrt{2\pi}a_0}}\sqrt{n_0}\simeq0.086.
\end{equation}
This justifies our choice of the range of values for $nU$ in the tight-binding model (Section \ref{appen:model}).

\subsection{Numerical simulations : truncated Wigner}
\label{appen:TW}

Numerical simulations were performed using the Truncated Wigner method~\cite{TW1,TW2,TW3} which allows one to account for the effect of quantum fluctuations. This method was implemented on the basis of a multiband description of the lattice problem at hand, using the Wannier orbitals $\chi_{n, \ell}(x) = \chi_{n,0}(x - \ell d)$ that are obtained from the inverse Fourier transform of the Bloch eigenstates of the homogeneous one-dimensional lattice described by the Hamiltonian:
\begin{equation}
  H_0 = \frac{\hat{p}^2}{2m} - \frac{s_0}{2} E_\dL \cos(k_\dL x)
\end{equation}
with $\hat{p} = - i \hbar \partial/\partial x$. Here, $\ell \in \mathbb{Z}$ is the lattice site index and $n$ represents the band index ranging between $0$, corresponding to the ground band, and a maximum excitation number $M$ chosen such that all relevant driving-induced intrawell coupling processes are accounted for in this representation. The Wannier orbitals are mutually orthogonal and normalized,
\begin{equation}
  \int_{-\infty}^\infty \chi_{n,\ell}^*(x) \chi_{n',\ell'}(x) dx = \delta_{n n'} \delta_{\ell \ell'}\,,
\end{equation}
and fulfill the parity property
\begin{equation}
  \chi_{n,0}(-x) = (-1)^n \chi_{n,0}(x) \label{eq:parity}
\end{equation}
owing to the symmetry of the lattice wells. On-site energies $E_n$ and nearest-neighbor hoppings $J_n$ associated with the $n^\mathrm{th}$ excited band are calculated from the relations
\begin{eqnarray}
  \int_{-\infty}^\infty \chi_{n,\ell}^*(x) H_0 \chi_{n',\ell}(x) dx & = & E_n \delta_{n n'} \,, \\
  \int_{-\infty}^\infty \chi_{n,\ell}^*(x) H_0 \chi_{n',\ell\pm 1}(x) dx & = & - J_n \delta_{n n'} \,,
\end{eqnarray}
respectively, while tunneling matrix elements beyond the nearest neighbors are neglected in the description.

We also neglect interaction effects involving Wannier orbitals on different sites, thus only accounting for on-site interaction matrix elements obtained from the integrals
\begin{equation}
  u_{n_1 n_2 n_1' n_2'} = \int_{-\infty}^\infty \chi_{n_1,\ell}^*(x) \chi_{n_2,\ell}^*(x)
  \chi_{n_1',\ell}(x) \chi_{n_2',\ell}(x) dx
\end{equation}
which, owing to the property \eqref{eq:parity}, vanish if $n_1 + n_2 + n_1' + n_2'$ is an odd number. Lattice shaking is incorporated through the gauge transformation
\begin{equation}
  \psi \mapsto \tilde{\psi} = \exp\left[-\frac{i \varphi_0}{\hbar k_\dL} \cos(2\pi\nu t) \hat{p} \right] \psi
\end{equation}
of the wavefunction, which effectively yields a periodically modulated synthetic gauge field. The associated matrix elements in the Wannier basis are given by
\begin{equation}
  p_{n n'}^{(\ell-\ell')} = \int_{-\infty}^\infty \chi_{n,\ell}^*(x) \hat{p} \chi_{n',\ell'}(x)
\end{equation}
and vanish for $\ell=\ell'$ if $n+n'$ is an even number.

\begin{figure}[ht!]
  \begin{center}
    \includegraphics[width=1\columnwidth]{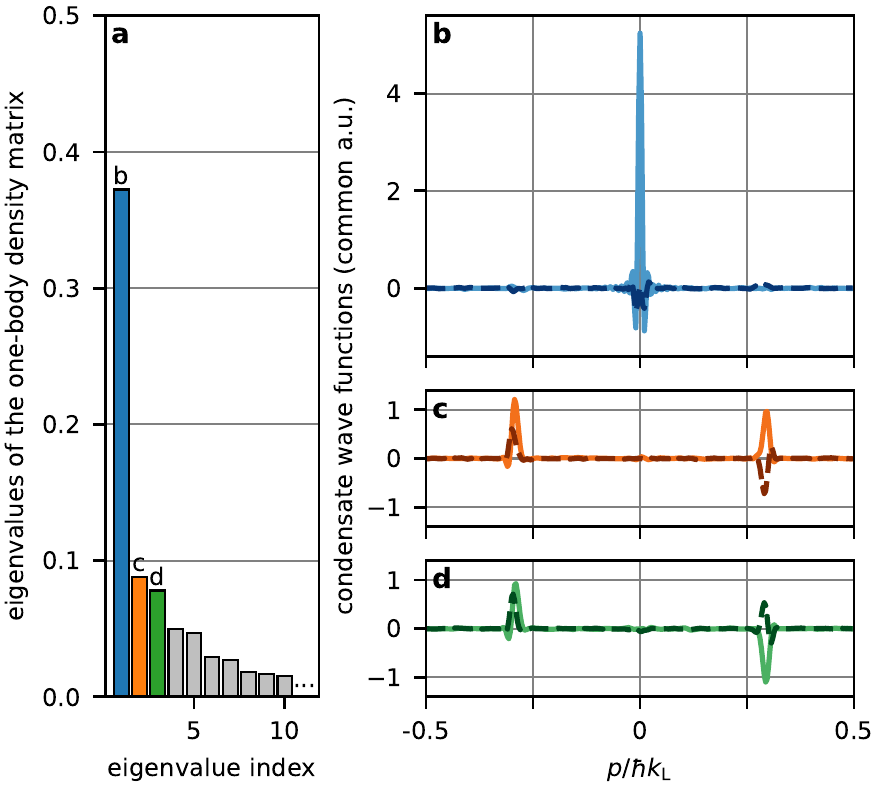}
    \caption{(a) Eigenvalues of the reduced one-body density matrix, normalized with respect to the total population of the atomic gas. Three eigenvalues are distinctly large with respect to the others, indicating the presence of Bose-Einstein condensates. (b-d) Ground-band components of the corresponding three associated condensate wavefunctions (obtained via the eigenvectors of the reduced one-body density matrix multiplied by the square roots of the associated eigenvalues) plotted in momentum space (solid line: real part; dashed line: imaginary part of the ground-mode wavefunction), in matching colors. Besides the primary condensate centered about $p=0$, two secondary condensates, corresponding to linear combinations of left- and right-moving states $e^{\pm i q^* x}$, are populated through four-wave mixing. Parameters: $s_0=3.4$, $\nu = 30$~kHz, $\varphi_0 = 20\degree$, $N=10^5$, $t = 5$~ms.}
    \label{appen:fig:TW2}
  \end{center}
\end{figure}

Neglecting driving-induced couplings beyond nearest neighbors, we obtain the time evolution equations for the calculations of trajectories in the framework of the quasi-classical Truncated Wigner method as
\begin{eqnarray}
  i\hbar \frac{\partial}{\partial t} \psi_{n,\ell}(t) & = & (E_n + V_\ell) \psi_{n,\ell}(t)
  - J_n [\psi_{n,\ell+1}(t) + \psi_{n,\ell-1}(t)] \nonumber \\
  && - \varphi_0\frac{2\pi\nu}{k_\dL} \sin(2\pi\nu t)
  \sum_{n'=0}^M \sum_{\ell'=\ell-1}^{\ell+1} p_{n n'}^{(\ell-\ell')} \psi_{n',\ell'}(t)
  \nonumber \\
  && + g_\ell \sum_{n', n_2, n_2'=0}^M u_{n n_2 n' n_2'}
  \psi_{n_2,\ell}^*(t) \psi_{n_2',\ell}(t) \psi_{n',\ell}(t) \nonumber \\
  && - g_\ell \sum_{n', n_2=0}^M u_{n n_2 n' n_2} \psi_{n',\ell}(t) \label{eq:TW}
\end{eqnarray}
with
\begin{equation}
  V_\ell = \frac{1}{2} m \omega_x^2 d^2 \ell^2
\end{equation}
the shift of the on-site energies due to the longitudinal confinement of the hybrid trap and
\begin{equation}
  g_\ell = (\sqrt{2\pi}/a_0) U_\ell = \frac{2\hbar\omega_\perp a_\mathrm{s}}{\sqrt{1+4n_\ell a_\mathrm{s}/(\sqrt{2\pi}/a_0)}}
\end{equation}
the effective on-site interaction parameter modified by the presence of the transverse confinement, as described in Sec.~\ref{appen:interaction}. The lattice site populations $n_\ell$ are numerically obtained from imaginary-time propagation yielding the initial condensate wavefunction in the un-driven lattice, and we assume here that they vary only marginally in the course of time evolution (which is not always the case, as seen in Fig.\ref{fig:figure4}). Note that the last term in Eq.~\eqref{eq:TW} arises from the proper derivation of the classical counterpart of the quantum interaction term via Weyl ordering.

The Truncated Wigner method allows one to compute quasi-classical expressions for the mean populations $\langle \hat{n}_{n,\ell} \rangle$ of the lattice sites as well as for the population correlations $\langle \hat{n}_{n,\ell} \hat{n}_{n',\ell'} \rangle$, where Weyl ordering has to be respected in order to correctly obtain the quantum expectation values from the classically calculated densities $|\psi_{n,l}(t)|^2$. It can also give access to the entire reduced one-body density matrix constituted by the expectation values of the coherence matrix elements $\langle \hat{a}_{n,\ell}^\dagger \hat{a}_{n',\ell'} \rangle$ (see Ref.~\cite{Wan21PRA} for a similar study). Diagonalization of this matrix yields the proper definition of the condensate fraction, via the eigenvector that is associated with its largest eigenvalue \cite{PenOns56PR}. We can thereby monitor the time evolution of the shape and population of the condensate, and the appearance and ultimate destruction of the non-commensurate crystalline order.

Within the time window where the emerging crystalline order is realized, the formation of two secondary Bose-Einstein condensates can be identified in the eigenspectrum of the reduced one-body density matrix, namely via the presence of two further eigenvalues that are distinctly large as compared to the rest of the spectrum. As displayed in Fig.\ref{appen:fig:TW2}, those two secondary condensate wavefunctions correspond to linear combinations of the two traveling waves $e^{\pm i q^* x}$ that are populated via four-wave-mixing. The superposition of those secondary condensates with the primary condensate centered about $p=0$ in momentum space gives rise to the coherence oscillations displayed in Fig.\ref{fig:figure4}.

\subsection{Experimental method : band-mapping}
\label{appen:bandmap}

\begin{figure*}
	\begin{center}
		\includegraphics[scale=1]{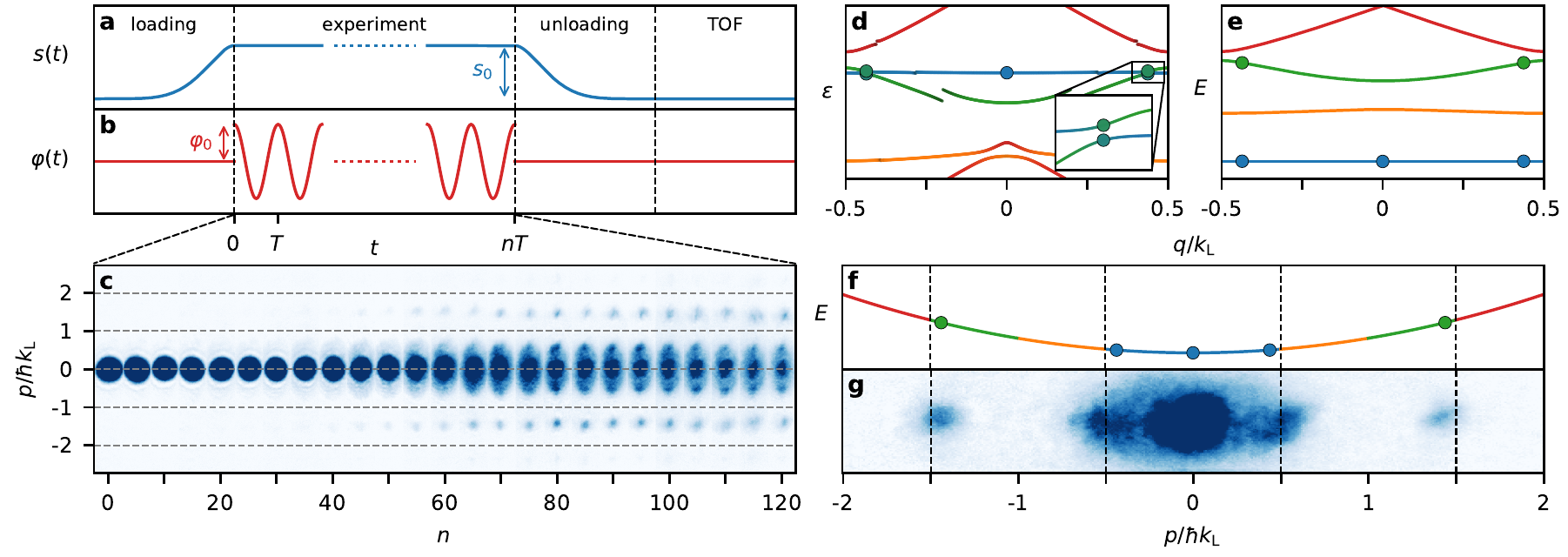}
		\caption{\textbf{Band-mapping process.} \textbf{(a)} Depth of the lattice along time: adiabatical loading at $s_0$, held constant during the experiment, adiabatical unloading to band map (see text) and switch off for time-of-flight imaging. \textbf{(b)} Phase of the lattice along time, sine-modulated with amplitude $\varphi_0$ for an integer number $n$ of periods $T$. \textbf{(c)} Stack of experimental absorption images for increasing $n$, with $s_0 = 3.70 \pm 0.10$, $\varphi_0= 15 \degree$, $\nu = 1/T = 25.5$ kHz and $t_\dTOF= 35$ ms. \textbf{(d)} Corresponding quasi-energy spectrum (colored lines) where the overlaps between the Floquet eigenstates and the eigenstates of the static lattice have been color-coded, with blue, orange, green and red corresponding respectively to the first 4 bands (\textit{s} to \textit{f}). BEC (disk in $q=0$) and instability (disks in $q\neq0$) modes. \textbf{(e-f)} Band structures of the lattices of depth $s_0=3.7$ for \textbf{(e)} and $s_0=0$ for \textbf{(f)} (solid colored lines) and follow-up of the modes (see text) with the same color code. \textbf{(f-g)} BZ borders (black dotted lines). \textbf{(g)} Absorption image after $n=80$ periods of data \textbf{(c)}.}
		\label{appen:fig:figure6}
	\end{center}
\end{figure*}

\begin{figure}
	\begin{center}
		\includegraphics[width=1\columnwidth]{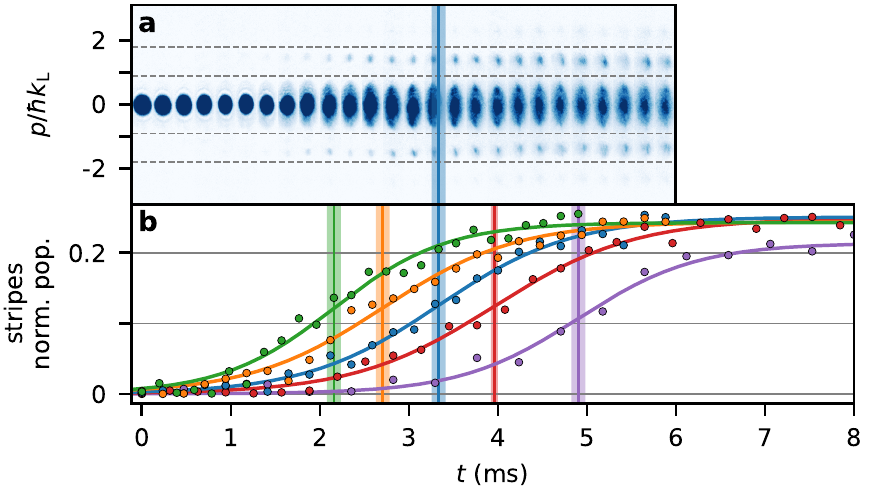}
		\caption{\textbf{Timescales measurements.} \textbf{(a)} Example of measured data series after band-mapping, for $s_0=3.70\pm 0.20$, $\nu=25.5\,\mathrm{kHz}$ and $\varphi_0=15^\circ$. The population from higher bands in the horizontal grey shaded stripes is measured over time to extract a nucleation time. \textbf{(b)} Growth curves extracted as in \textbf{(a)} for the points of Fig.\ref{fig:figure3}(b) corresponding to coupling bands $s$ and $d$, with the purple, red, blue, orange, and green data corresponding to $\varphi_0=\{10\,\degree,12.5\,\degree,15\,\degree,17.5\,\degree,20\,\degree\}$ respectively. The sigmoid fitting curves are shown and the extracted nucleation times are represented by vertical lines, with shaded areas denoting the uncertainty from the fit.}
		\label{appen:fig:figure7}
	\end{center}
\end{figure}

To more accurately identify the onset of the emerging crystalline order, we use the band-mapping technique \cite{bandmapping} that consists in decreasing adiabatically the lattice before the time-of-flight (see Fig.\ref{appen:fig:figure6}). The band-mapping reveals the hybrid nature of the unstable modes induced by resonant coupling. The clear growth of the population fraction in the higher coupled band is subsequently plotted as a function of time to characterize the kinetics of the crystalline state formation  and its persistence over time (see Fig.\ref{appen:fig:figure7}).

\end{appendix}

\end{document}